\documentclass[journal]{IEEEtran}
\IEEEoverridecommandlockouts

\usepackage{cite}
\usepackage{amsmath,amssymb,amsfonts}
\usepackage{graphicx}
\usepackage{multirow}
\usepackage{subcaption}
\usepackage{caption}
\usepackage{float}
\usepackage{textcomp}
\usepackage{xcolor}
\usepackage{tikz}
\usepackage{pgfplots}
\usepackage{stfloats}
\usepackage{mathrsfs}
\usepackage{balance}
\usepackage{amsthm}
\usepackage{wasysym} 
\usepackage[ruled,vlined]{algorithm2e}

\newtheorem{remark}{Remark}
\usepackage{comment}
 \usepackage{booktabs}
\usepackage{algpseudocode}
\usepackage{acronym}
\usepackage[hyperfirst=true]{glossaries}
\usepackage[hidelinks]{hyperref} 
\usepackage{makecell}
\makeglossaries
\newacronym{OFDM}{OFDM}{orthogonal frequency division multiplexing}
\newacronym{OTFS}{OTFS}{orthogonal time frequency space}
\newacronym{AFDM}{AFDM}{affine frequency division multiplexing}
\newacronym{SIMO}{SIMO}{single-input multiple-output}
\newacronym{DL}{DL}{deep learning}
\newacronym{EVM}{EVM}{error vector magnitude}
\newacronym{BER}{BER}{bit error rate}
\newacronym{HSR}{HSR}{high-speed railways}
\newacronym{V2V}{V2V}{vehicle-to-vehicle}
\newacronym{V2I}{V2I}{vehicle-to-infrastructure}
\newacronym{V2X}{V2X}{vehicle-to-everything}
\newacronym{UAV}{UAV}{unmanned aerial vehicle}
\newacronym{ICI}{ICI}{intercarrier interference}
\newacronym{ISI}{ISI}{intersymbol interference}
\newacronym{IPI}{IPI}{interpath interference}
\newacronym{CFR}{CFR}{channel frequency response}
\newacronym{SISO}{SISO}{single-input single-output}
\newacronym{DoA}{DoA}{direction-of-arrival}
\newacronym{TX}{TX}{transmitter}
\newacronym{RX}{RX}{receiver}
\newacronym{CP}{CP}{cyclic prefix}
\newacronym{RMSE}{RMSE}{root mean squared error}
\newacronym{CRLB}{CRLB}{Cramér-Rao Lower Bound}
\newacronym{AWGN}{AWGN}{additive white gaussian noise}
\newacronym{SNR}{SNR}{signal-to-noise ratio}
\newacronym{MSE}{MSE}{mean squared error}
\newacronym{MLP}{MLP}{multilayer perceptron}
\newacronym{FI}{FI}{Fisher information}
\newacronym{DD}{DD}{decision-directed}
\newacronym{ULA}{ULA}{uniform linear array}
\newacronym{DFT}{DFT}{discrete Fourier transform}
\newacronym{QAM}{QAM}{quadrature amplitude modulation}
\newacronym{MF}{MF}{matched filter}
\newacronym{MRC}{MRC}{maximum ratio combining}
\newacronym{CFAR}{CFAR}{constant false alarm rate}
\newacronym{LS}{LS}{least squares}
\newacronym{FNN}{FNN}{feedforward neural network}
\newacronym{CSI}{CSI}{channel state information}
\newacronym{zD}{zD}{zero Doppler}
\newacronym{PSD}{PSD}{power spectral density}
\newacronym{ReLU}{ReLU}{rectified linear unit}
\newacronym{MMSE}{MMSE}{minimum mean squared error}
\newacronym{CE}{CE}{channel estimation}
\newacronym{PDR}{PDR}{pilot-to-data ratio}
\newacronym{EP}{EP}{embedded-pilot}
\newacronym{SP}{SP}{superimposed pilot}
\newacronym{RE}{RE}{resource element}
\newacronym{TM}{TM}{threshold method}
\newacronym{IMFC}{IMFC}{iterative matched filtering and combining}
\newacronym{ISFFT}{ISFFT}{inverse symplectic finite Fourier trasform}
\newacronym{PAPR}{PAPR}{peak-to-average power ratio}
\newacronym{DD-a-OFDM}{DD-a-OFDM}{delay-Doppler aided OFDM}
\newacronym{ISAC}{ISAC}{integrated sensing and communication}
\newacronym{PA}{PA}{power amplifiers}

\usepackage{orcidlink}

\begin{document}

\title{Robust 6G OFDM High-Mobility Communications Using Delay-Doppler Superimposed Pilots}

\author{
Mauro Marchese\orcidlink{0009-0008-0265-5840},~\IEEEmembership{Graduate Student Member,~IEEE}, Pietro Savazzi\orcidlink{0000-0003-0692-8566},~\IEEEmembership{Senior Member,~IEEE}%
\thanks{M. Marchese is with the Department of Electrical, Computer and Biomedical Engineering, University of Pavia, Pavia, 27100 Italy (e-mail: mauro.marchese01@universitadipavia.it).}
\thanks{Pietro Savazzi is with the Department of Electrical, Computer and Biomedical Engineering, University of Pavia, Pavia, 27100 Italy (e-mail: pietro.savazzi@unipv.it), and with the Consorzio Nazionale Interuniversitario per le Telecomunicazioni - CNIT.}
}

\maketitle

\begin{abstract}
In this work, a novel receiver architecture for \gls{OFDM} communications in 6G high-mobility scenarios is developed. In particular, a delay-Doppler \gls{SP} scheme is used for \gls{CE} by adding a single pilot in the delay-Doppler domain. Unlike previous research on delay-Doppler superimposed pilots in \gls{OFDM} systems,  \gls{ICI} effects, fractional delays, and Doppler shifts are considered. Consequently, a disjoint fractional delay-Doppler estimation algorithm is derived, and a reduced-complexity equalization method based on the Landweber iteration, which exploits intrinsic channel structure, is proposed. Simulation results reveal that the proposed receiver architecture achieves robust communication performance across various mobility conditions, with speeds of up to 1000 km/h, and increases the effective throughput compared to existing methods.
\end{abstract}


\glsresetall

\section{Introduction}
\IEEEPARstart{T}{oday}, wireless communication systems widely adopt the \gls{OFDM} waveform to achieve high spectral efficiency and robust performance. The attractiveness of \gls{OFDM} stems from its resilience to \gls{ISI} in multipath environments \cite{Litwin2001, Viterbo2022}. In particular, the use of the \gls{CP} guarantees \gls{ISI}-free operation, provided that the \gls{CP} length is designed to be greater than the channel delay spread. As a consequence, subcarrier orthogonality remains intact, and the per-subcarrier channels can be modeled as \gls{AWGN} channels.
However, this optimal performance is experienced only in static multipath channels. When mobility occurs, the channel becomes time-varying \cite{Viterbo2022}. Under low-mobility conditions, the \gls{CFR} changes slowly over time and can be considered constant within one \gls{OFDM} symbol. In such cases, \gls{OFDM} can achieve near-optimal performance, as the time-frequency \glspl{RE} remain almost orthogonal at the receiver. Therefore, the simple and well-known single-tap equalizer can be adopted at the receiver \cite{Viterbo2022,Bello2025}.

In recent years, high-mobility communications have become increasingly important for the development of next-generation wireless systems, such as 6G \gls{V2X}, \gls{HSR}, and \glspl{UAV}. In these scenarios, speeds of up to $1000$~km/h \cite{Giordani2020,Tataria2021} can be experienced, and Doppler effects become prominent. Under such conditions, orthogonality is destroyed by \gls{ICI} \cite{Robertson1999,Robertson1999_2,Mostofi2005,Wang2006}, and the performance of \gls{OFDM} degrades. Moreover, the advantage of low-complexity symbol detection via single-tap equalization vanishes, as more complex equalization methods, such as \gls{MMSE} \cite{Choi2001,Yiyan2025}, must be adopted. As a consequence, novel waveforms have been introduced to overcome the limitations of \gls{OFDM} \cite{Zhou2024}. In particular, a waveform known as \gls{OTFS} was proposed in \cite{Hadani2017} and has garnered significant attention due to its performance gains over \gls{OFDM}. \gls{OTFS} operates in the delay-Doppler domain \cite{Hadani2017,Mohammed2022,Viterbo2022}, where the channel can be considered sparse and nearly time-invariant \cite{Hadani2017,Viterbo2022}.

Accordingly, several works have addressed \gls{OTFS} \gls{CE} in the delay-Doppler domain by using either dedicated pilots \cite{Raviteja2019,Khan2021,Khan2023,marchese2024,Marchese2025disjoint} or \glspl{SP} \cite{Mishra2022,kanazawa2025}. In \gls{OTFS}, the \gls{CE} problem is formulated as a parameter estimation problem, where the objective is to estimate the channel parameters, namely the channel gains, delays, and Doppler shifts.

On the other hand, conventional \gls{CE} schemes in \gls{OFDM} systems rely on \glspl{EP} in the time-frequency domain and directly estimate the channel response at pilot positions, followed by interpolation to recover the full channel response \cite{Colieri2002,Colieri2002_2}. \gls{SP}-based schemes for \gls{OFDM} systems have also been investigated in \cite{Cui2005,Varma2006,Li2006,Lu2007,Josiam2007,He2012,Zhang2014}. In these works, frequency-domain pilots are linearly superimposed onto data symbols across subcarriers.

More recently, the idea of incorporating \gls{OTFS}-like delay-Doppler processing into \gls{OFDM} systems has attracted increasing attention, as it allows one to exploit the channel properties in the delay-Doppler domain, namely sparsity and the large geometric coherence time of the channel parameters (channel gains, propagation delays, and Doppler shifts) \cite{Viterbo2022}. In \cite{Shaw2023,Shaw2025}, parameter-based channel estimation is performed by transmitting dedicated pilots in the time-frequency domain, achieving more robust communication performance. Similarly, in \cite{Yiyan2025}, a \gls{DD-a-OFDM} scheme is proposed for high-mobility scenarios by exploiting \gls{CE} in the delay-Doppler domain together with conventional \gls{MMSE} equalization in the time-frequency domain.

In \cite{Bello2025}, a novel pilot scheme for \gls{OFDM} that exploits the superposition of pilots in the delay-Doppler domain is proposed. Specifically, a single delay-Doppler pilot is superimposed onto the time-frequency \gls{OFDM} symbol matrix, and channel estimation is performed by searching for peaks in the delay-Doppler domain, in a manner similar to \gls{OTFS}-based \gls{SP} channel estimation \cite{Mishra2022,kanazawa2025}. This approach is sometimes referred to as \gls{TM} in the \gls{OTFS} literature \cite{Viterbo2022,marchese2024,Marchese2025disjoint}. However, the work in \cite{Bello2025} relies on two simplifying assumptions: (i) \gls{ICI}-free operation, such that the channel is constant within an \gls{OFDM} symbol and single-tap equalization can be employed at the receiver; and (ii) integer delays and Doppler shifts, implying that the channel parameters are multiples of the delay and Doppler resolutions. In practical 6G high-mobility scenarios, these assumptions do not generally hold. Therefore, the following question arises: \textit{is it possible to achieve robust communication performance in 6G \gls{OFDM} high-mobility scenarios in the presence of \gls{ICI} and fractional channel parameters by leveraging delay-Doppler \gls{SP} schemes?} This work aims to address this question by: (i) relaxing the integer delay-Doppler assumption in \cite{Bello2025} and developing a novel delay-Doppler \gls{SP}-based \gls{CE} algorithm for \gls{OFDM}; and (ii) introducing an \gls{ICI}-aware equalization scheme for \gls{OFDM} with lower computational complexity than the conventional full-\gls{MMSE} detector \cite{Choi2001,Yiyan2025}.

\subsection{Contributions}
Given the above discussion, the contributions of this work are summarized as follows:
\begin{itemize}
    \item \textbf{Novel fractional delay-Doppler estimation based on a \gls{SP} scheme}: By relaxing the oversimplifying assumptions made in \cite{Bello2025}, a novel algorithm for disjoint estimation of fractional delays and Doppler shifts is developed. In particular, under the proposed delay-Doppler superimposed pilot scheme, the integer components are obtained by searching for the maximum pilot energy in the delay-Doppler domain. Subsequently, fractional delays and Doppler shifts are estimated by correlating the received profiles with the corresponding delay and Doppler terms. This approach enhances \gls{CE} performance in realistic scenarios with fractional channel parameters, while maintaining the low computational complexity inherited from the disjoint estimation process.
    
    \item \textbf{Low-complexity equalization based on the Landweber method}: A path-wise low-complexity equalizer, termed \gls{IMFC}, based on the Landweber algorithm \cite{CharlesByrne_2004,Hung2007}, is developed for symbol detection. Specifically, the intrinsic structure of the channel is exploited to decompose the iterative algorithm into multiple low-complexity operations. The computational complexity of this approach scales linearly with the number of time-frequency \glspl{RE}, in contrast to the conventional \gls{MMSE} equalizer, whose complexity scales cubically with the number of subcarriers \cite{Choi2001,Yiyan2025}.
    
    \item \textbf{Mobility-resilient numerical analysis under \gls{ICI} and fractional delays and Doppler shifts}: Extensive simulations are conducted to evaluate the robustness of the proposed delay-Doppler \gls{SP}-based \gls{OFDM} receiver against mobility in practical scenarios characterized by \gls{ICI} and fractional channel parameters. Different modulation formats, \glspl{SNR}, and user speeds are considered. The numerical results demonstrate the robustness of the proposed \gls{OFDM} receiver against mobility. In particular, the proposed scheme achieves an approximately constant effective throughput over a wide range of maximum speeds, up to $1000$~km/h, which corresponds to the maximum velocity targeted by 6G networks \cite{Giordani2020,Tataria2021}.
\end{itemize}

\textit{Notation}: $x$ is a scalar, $\mathbf{x}$ is a vector, and $\mathbf{X}$ is a matrix. The $(m,n)$-th element of matrix $\mathbf{X}$ is denoted as $[\mathbf{X}]_{m,n}$. For a matrix $\mathbf{X}$, $[\mathbf{X}]_{:,n}$ and $[\mathbf{X}]_{m,:}$ denote the $n$-th column and the $m$-th row of $\mathbf{X}$, respectively. $\mathbf{X}^*$, $\mathbf{X}^\top$, $\mathbf{X}^{\mathsf{H}}$ and $\mathbf{X}^{-1}$ represent the complex conjugate, transpose, Hermitian (conjugate transpose) and inverse of $\mathbf{X}$. Moreover,
$\|\mathbf{X}\|_F$ and $\text{Tr}(\mathbf{X})$  represent the Frobenius norm and the trace of $\mathbf{X}$, respectively. The operator $\text{diag}(\mathbf{x})$ denotes the diagonal matrix whose main diagonal is given by the elements of vector $\mathbf{x}$. $\mathbf{I}_N$ denotes the identity matrix of order $N$. The notation $[\cdot]_{N}$ indicates the modulo-$N$ operation. $\odot$ and $\circledast$ represent Hadamard (element-wise) and 2D circular convolution, respectively. The expectation operator is denoted as $\mathbb{E}[\cdot]$. The distribution $\mathcal{CN}(\boldsymbol{\mu},\mathbf{\Sigma})$ denotes a circularly symmetric complex Gaussian random vector with mean $\boldsymbol{\mu}$ and covariance matrix $\mathbf{\Sigma}$. Finally, $\delta_{ij}$ is the Kronecker delta that takes the value $1$ when $i=j$ and zero otherwise.

\section{System Model}\label{sec:SysModel}
In this section, the \gls{SISO} \gls{OFDM} system model is presented, and both \gls{ICI}-aware and \gls{ICI}-free models are discussed.

\subsection{Continuous-Time Model}

\subsubsection{OFDM Transmit Signal Model}
The \gls{TX} arranges $M \times N$ symbols in the time-frequency plane across $M$ subcarriers with spacing $\Delta f$ and $N$ time slots. Accordingly, the transmit symbol matrix $\mathbf{X} \in \mathbb{C}^{M \times N}$ is obtained. Denoting $[\mathbf{X}]_{m,n} = x_{m,n}$ as the $(m,n)$-th \gls{RE}, the transmit signal during the $n$-th symbol duration is given by \cite{Keskin2021}
\begin{equation}\label{eq:txsignalmodel}
s_n(t) = \frac{1}{\sqrt{M}} \sum_{m=0}^{M-1} x_{m,n} e^{j2\pi m \Delta f t} \Pi\Bigg( \frac{t - nT'}{T'} \Bigg),
\end{equation}
where $\Pi(t)$ is a rectangular pulse that takes the value $1$ for $t \in [0,1]$ and $0$ otherwise, and $T'$ denotes the overall \gls{OFDM} symbol duration. Specifically, $T' = T + T_{\text{CP}}$, where $T = 1/\Delta f$ and $T_{\text{CP}} > \sigma_\tau$ denotes the duration of the \gls{CP}, with $\sigma_\tau$ being the channel delay spread. Therefore, as in standard \gls{OFDM} transmission, \gls{ISI} is avoided. In the following, the overall signal including the \gls{CP} is denoted by $s_n^{\text{CP}}(t)$.

\subsubsection{Channel Model}
The high-mobility wireless channel consists of $P$ propagation paths, each characterized by a delay $\tau_p$, a Doppler shift $\nu_p$, and a complex channel gain $\alpha_p$. The time-varying impulse response of the \gls{SISO} channel is given by
\begin{equation}
h(t,\tau) = \sum_{p=1}^P \alpha_p e^{j2\pi \nu_p t} \delta(\tau - \tau_p).
\end{equation}

\subsubsection{Received Signal Model}
During the $n$-th symbol interval, the received signal is expressed as
\begin{equation}
\begin{split}\label{eq:RxMultipathCP}
r_n^{\text{CP}}(t) &= \int h(t,\tau) s_n^{\text{CP}}(t - \tau) \, \mathrm{d}\tau + n(t) \\
&= \sum_{p=1}^P \alpha_p s_n^{\text{CP}}(t - \tau_p) e^{j2\pi \nu_p t} + n(t),
\end{split}
\end{equation}
where $n(t)$ denotes \gls{AWGN} with one-sided \gls{PSD} $N_0$.

\subsection{ICI-Aware Model in Matrix Form}
By sampling \eqref{eq:txsignalmodel} at the symbol rate $T/M$, the transmit signal vector is given by
\begin{equation}\label{eq:txsignalvector}
\mathbf{s}_n = \mathbf{F}^{\mathsf{H}}_M \mathbf{x}_n \in \mathbb{C}^{M \times 1},
\end{equation}
where $\big[\mathbf{F}_{M}\big]_{m,q} = \frac{1}{\sqrt{M}} e^{-j2\pi \frac{mq}{M}}$ denotes the $M$-point \gls{DFT} matrix. Moreover, the average transmit power is defined as $P_T = \mathbb{E}[\|\mathbf{S}\|_F^2]/(MN)$, where $\mathbf{S} = [\mathbf{s}_0 \ \mathbf{s}_1 \ \dots \ \mathbf{s}_{N-1}] \in \mathbb{C}^{M \times N}$ denotes the transmit signal matrix.

Similarly, by sampling \eqref{eq:RxMultipathCP} at the same rate and removing the \gls{CP}, the received observation vectors $\mathbf{r}_n \in \mathbb{C}^{M \times 1}$ are obtained as \cite{Keskin2021,Keskin2024}
\begin{equation}\label{eq:timespatialobs}
\mathbf{r}_n = \mathbf{H}_n \mathbf{s}_n + \mathbf{n}_n,
\end{equation}
where $\mathbf{n}_n \sim \mathcal{CN}(\mathbf{0}, \sigma^2 \mathbf{I}_{M})$ denotes the \gls{AWGN} vector, and $\mathbf{H}_n \in \mathbb{C}^{M \times M}$ is the channel matrix given by
\begin{equation}\label{eq:delaytimeChMatrix}
\mathbf{H}_n = \sum_{p=1}^P \tilde{\alpha}_{p,n} \, \mathbf{\tilde{C}}(\nu_p)\mathbf{F}_{M}^{\mathsf{H}}\mathbf{B}(\tau_p)\mathbf{F}_{M}.
\end{equation}
Here, $\mathbf{B}(\tau) = \mathrm{diag}(\mathbf{b}(\tau))$, with $\mathbf{b}(\tau) = \big[e^{-j2\pi q \tau \Delta f}\big]_{q=0}^{M-1}$, and $\mathbf{\tilde{C}}(\nu) = \mathrm{diag}(\mathbf{\tilde{c}}(\nu))$, where $\mathbf{\tilde{c}}(\nu) = \big[e^{j2\pi q \nu \frac{T}{M}}\big]_{q=0}^{M-1}$ captures fast-time variations due to Doppler, i.e., \gls{ICI}. Furthermore, $\tilde{\alpha}_{p,n} = \alpha_p e^{j2\pi \nu_p t_n}$ denotes the Doppler-induced time-varying channel gain, where $t_n = T_{\text{CP}} + nT'$ is the time instant at which the transmission of the $n$-th \gls{OFDM} symbol starts. This term captures the slow-time variations induced by Doppler.

By concatenating the received observation vectors as $\mathbf{R} = [\mathbf{r}_0 \ \mathbf{r}_1 \ \dots \ \mathbf{r}_{N-1}]$, the received \gls{OFDM} frame can be expressed as \cite{Keskin2021}
\begin{equation}\label{eq:delaytimeObservations}
\mathbf{R} = \sum_{p=1}^P \alpha_p \mathbf{\tilde{C}}(\nu_p)\mathbf{F}_{M}^{\mathsf{H}}\Big( \mathbf{X} \odot \mathbf{b}(\tau_p)\mathbf{c}^\top(\nu_p) \Big) + \mathbf{N} \in \mathbb{C}^{M \times N},
\end{equation}
where $\mathbf{c}(\nu) \in \mathbb{C}^{N \times 1}$ is the slow-time Doppler steering vector defined as $\mathbf{c}(\nu) = \big[e^{j2\pi \nu t_n}\big]_{n=0}^{N-1}$, and $\mathbf{N} = [\mathbf{n}_0 \ \mathbf{n}_1 \ \dots \ \mathbf{n}_{N-1}]$ denotes the noise matrix.

The input--output relation in \eqref{eq:delaytimeObservations} can be compactly rewritten as
\begin{equation}
\mathbf{R} = \boldsymbol{\mathcal{H}}(\mathbf{X}) + \mathbf{N},
\end{equation}
where $\boldsymbol{\mathcal{H}}(\cdot) = \sum_{p=1}^P \alpha_p \Big[ \mathbf{F}_{M}^{\mathsf{H}}\big( (\cdot) \odot \mathbf{H}_{\mathrm{tf}}^{(p)} \big) \odot \mathbf{H}_{\mathrm{ICI}}^{(p)} \Big]$ denotes the channel operator. Moreover, $\mathbf{H}_{\mathrm{tf}}^{(p)}$ is the frequency--time channel matrix associated with the $p$-th propagation path, given by
\begin{equation}
\mathbf{H}_{\mathrm{tf}}^{(p)} = \mathbf{b}(\tau_p)\mathbf{c}^{\top}(\nu_p),
\end{equation}
and $\mathbf{H}_{\mathrm{ICI}}^{(p)}$ is the delay--time \gls{ICI} matrix accounting for the \gls{ICI} generated by the $p$-th path, defined as
\begin{equation}
\mathbf{H}_{\mathrm{ICI}}^{(p)} = \mathbf{\tilde{c}}(\nu_p)\mathbf{1}_N^\top.
\end{equation}

\subsection{ICI-Free Model in Matrix Form}
The model in \eqref{eq:delaytimeObservations} simplifies when the subcarrier spacing satisfies $\Delta f \gg \nu_{\max}$, where $\nu_{\max} = \frac{v_{\max}}{c} f_c$, and $v_{\max}$ and $f_c$ denote the maximum speed and the carrier frequency, respectively. Under this condition, \gls{ICI} becomes negligible, such that $\mathbf{\tilde{C}}(\nu) \approx \mathbf{I}_M$, and the well-known \gls{ICI}-free model is obtained \cite{Bello2025}
\begin{equation}\label{eq:delaytimeObservations_ICIfree}
\mathbf{R} = \mathbf{F}_{M}^{\mathsf{H}} \sum_{p=1}^P \alpha_p \Big( \mathbf{X} \odot \mathbf{b}(\tau_p)\mathbf{c}^\top(\nu_p) \Big) + \mathbf{N}.
\end{equation}
Moreover, by transforming the delay--time observations in \eqref{eq:delaytimeObservations_ICIfree} to the frequency--time domain, one obtains
\begin{equation}\label{eq:ICIfreeModel}
\mathbf{Y} = \mathbf{F}_M \mathbf{R} = \mathbf{H}_{\mathrm{tf}} \odot \mathbf{X} + \mathbf{N},
\end{equation}
where the frequency--time channel matrix is given by \cite{Bello2025}
\begin{equation}\label{eq:TFchannel_ICIfree}
\mathbf{H}_{\mathrm{tf}} = \sum_{p=1}^P \alpha_p \mathbf{b}(\tau_p)\mathbf{c}^\top(\nu_p).
\end{equation}
In this case, the receiver can employ a single-tap equalizer to compensate for the channel effects.

\section{Channel Estimation with Delay-Doppler Superimposed Pilots}
This section first presents the conventional \gls{EP} method for channel estimation, followed by the delay-Doppler superimposed pilot model proposed in \cite{Bello2025} for \gls{CE} in \gls{OFDM} systems. Subsequently, the threshold method used in \cite{Bello2025} is reviewed, and its limitations are discussed. Finally, the proposed iterative \gls{CE} algorithm accounting for fractional channel parameters is presented.

\subsection{Conventional Channel Estimation in the Time-Frequency Domain using Embedded Pilots}
\label{sec:dedicated_pilot_ft}
The conventional channel estimation scheme in \gls{OFDM} systems assumes negligible \gls{ICI} and therefore relies on the \gls{ICI}-free model in \eqref{eq:ICIfreeModel}. Channel estimation is based on dedicated pilots in the time-frequency domain, commonly referred to as the \gls{EP} scheme. In this method, pilot symbols are allocated to specific \glspl{RE} to maximize the \gls{SNR} and minimize pilot-data interference. The channel is first estimated at pilot locations and then reconstructed over the entire time-frequency grid using interpolation techniques applied to \eqref{eq:TFchannel_ICIfree}.

Specifically, pilots are inserted into the time-frequency grid according to a defined periodicity along both frequency and time, resulting in a non-zero pilot density $\mathcal{D}_p$ \cite{Bello2025}. The pilot density is defined as
\begin{equation}
\mathcal{D}_p = \frac{1}{K_t K_f},
\end{equation}
where $K_f$ and $K_t$ denote the spacings between pilots along the frequency and time axes, respectively. The transmitted signal matrix $\mathbf{X}$ contains pilot symbols in dedicated positions $\mathcal{X}_p$, which carry known information, and data symbols (e.g., \gls{QAM}) occupying the remaining \glspl{RE} $\mathcal{X}_d$. Accordingly, $\mathbf{X}$ is constructed as
\begin{equation}
 x_{m,n} = 
\begin{cases}
\text{pilot}, & \forall (m,n) \in \mathcal{X}_p,\\
\text{data}, & \forall (m,n) \in \mathcal{X}_d.
\end{cases}
\end{equation}
Channel estimation in the dedicated pilot scheme is typically performed in two steps: first, estimation at pilot locations, and second, interpolation across the data \glspl{RE}.

\subsubsection{Channel Estimation at Pilot Locations}
The channel response is estimated only at the pilot locations $(m,n) \in \mathcal{X}_p$. Given the received time-frequency signal in \eqref{eq:ICIfreeModel}, the \gls{LS} estimate of the channel at pilot locations is given by
\begin{equation}
[\hat{\mathbf{H}}_{\mathrm{tf}}]_{m,n} = \frac{[\mathbf{Y}]_{m,n}}{x_{m,n}}, \quad \forall (m,n) \in \mathcal{X}_p.
\end{equation}

\subsubsection{Channel Interpolation}
To obtain the complete \gls{CSI} estimate $\hat{\mathbf{H}}_{\mathrm{tf}}$ over the entire $M \times N$ grid, interpolation is required. Sequential linear interpolation is typically performed in two steps: first along one dimension (e.g., frequency) and then along the other dimension (time) using the results from the first step.

For a data \gls{RE} $(m',n)$ located between two adjacent pilots in the frequency domain at $m_p$ and $m_p + K_f$, the interpolated estimate is
\begin{equation}
\begin{split}
[\hat{\mathbf{H}}_{\mathrm{tf}}]_{m',n} &= [\hat{\mathbf{H}}_{\mathrm{tf}}]_{m_p,n} \\
&\quad + \frac{m' - m_p}{K_f} \Big( [\hat{\mathbf{H}}_{\mathrm{tf}}]_{m_p+K_f,n} - [\hat{\mathbf{H}}_{\mathrm{tf}}]_{m_p,n} \Big).
\end{split}
\end{equation}
After obtaining estimates across all subcarriers $m'$ (from $0$ to $M-1$), the interpolation proceeds along the time dimension. For a data \gls{RE} $(m',n')$ located between two adjacent pilots in the time domain at $n_p$ and $n_p + K_t$, the interpolation uses the estimates obtained from the previous step:
\begin{equation}
\begin{split}
[\hat{\mathbf{H}}_{\mathrm{tf}}]_{m',n'} &= [\hat{\mathbf{H}}_{\mathrm{tf}}]_{m',n_p} \\
&\quad + \frac{n' - n_p}{K_t} \Big( [\hat{\mathbf{H}}_{\mathrm{tf}}]_{m',n_p+K_t} - [\hat{\mathbf{H}}_{\mathrm{tf}}]_{m',n_p} \Big).
\end{split}
\end{equation}
This procedure is applied sequentially along both dimensions to reconstruct the full $\hat{\mathbf{H}}_{\mathrm{tf}}$ matrix. 

The performance of the conventional \gls{EP} scheme is limited by the Doppler-induced time-varying nature of the channel. In particular, in high-mobility scenarios, \gls{ICI} destroys subcarrier orthogonality, causing the \gls{ICI}-free \gls{EP} scheme to fail.

\begin{figure}[t]
\centering
    \includegraphics[width=\columnwidth]{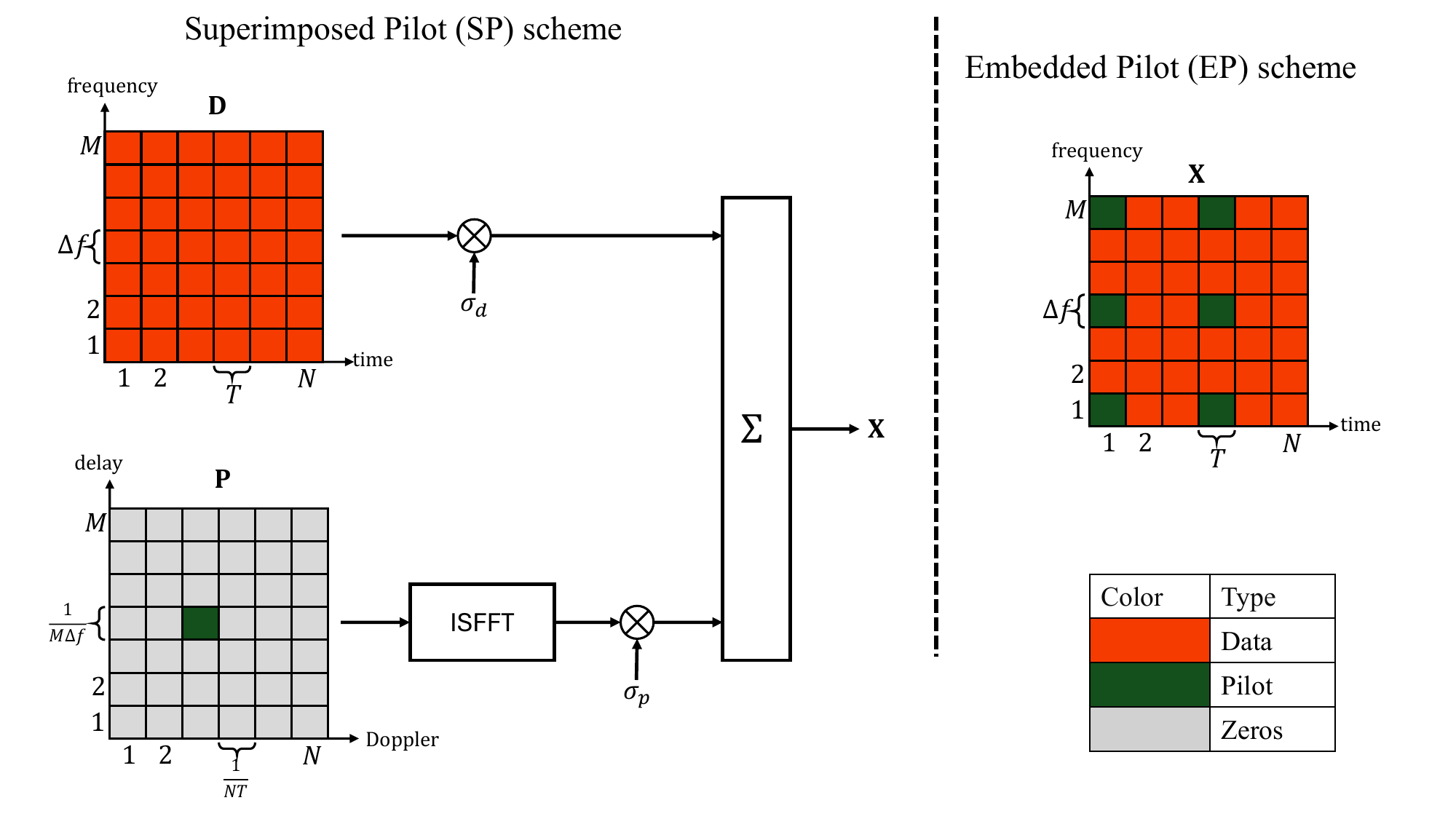}
    \caption{The delay-Doppler superimposed pilot (SP) scheme and the conventional time-frequency embdedded pilot (EP) scheme adopted in \gls{OFDM} systems.}\label{fig:PilotSchemes}
\end{figure}

\subsection{Delay-Doppler Pilot Superimposition}\label{sec:SPscheme}
In this section, the delay-Doppler \gls{SP} scheme proposed in \cite{Bello2025} is introduced. Figure \ref{fig:PilotSchemes} illustrates both the conventional \gls{EP} approach and the delay-Doppler \gls{SP} scheme considered in \cite{Bello2025}.

Let $\mathbf{P} \in \mathbb{C}^{M \times N}$ denote the delay-Doppler pilot matrix. A unit-amplitude pilot is placed for channel estimation as \cite{Bello2025}
\begin{align}\label{eq:PilotModel_elem}
[\mathbf{P}]_{m,n} = \delta_{mm_p}\delta_{nn_p} =
\begin{cases}
1, & m = m_p, ~ n = n_p, \\
0, & \text{otherwise},
\end{cases}
\end{align}
which can equivalently be expressed as
\begin{equation}\label{eq:PilotMtx}
\mathbf{P} = \mathbf{e}_{m_p} \mathbf{e}_{n_p}^\top,
\end{equation}
where $[\mathbf{e}_i]_j = \delta_{ij}$ denotes a one-hot vector with a single one at position $i$. It follows that $\|\mathbf{P}\|_F^2 = 1$.

This delay-Doppler pilot is superimposed onto \gls{OFDM} frequency-time data via the \gls{ISFFT} as
\begin{equation}\label{eq:SuperimposedPilotModel}
\mathbf{X} = \sigma_d \mathbf{D} + \sigma_p \underbrace{\mathbf{F}_M \mathbf{P} \mathbf{F}_N^{\mathsf{H}}}_{\text{ISFFT}} \in \mathbb{C}^{M \times N},
\end{equation}
where $\mathbf{D}$ contains data symbols (e.g., \gls{QAM}) with $\mathbb{E}\big[ |d_{m,n}|^2 \big] = 1$, where $d_{m,n} = [\mathbf{D}]_{m,n}$, so that $\mathbb{E}\big[ \|\mathbf{D}\|_F^2 \big] = MN$. Here, $\sigma_d^2$ and $\sigma_p^2$ denote the power allocated to data and pilot symbols, respectively.

Defining the channel gains vector as $\boldsymbol{\alpha} = [\alpha_1, \alpha_2, \dots, \alpha_P]^\top$ and assuming $\|\boldsymbol{\alpha}\|^2 = 1$, the \gls{SNR} is given by
\begin{equation}
\begin{split}
\mathrm{SNR} &= \frac{P_T}{\sigma^2} = \frac{\mathbb{E}[\|\mathbf{X}\|_F^2]}{\mathbb{E}[\|\mathbf{N}\|_F^2]} \\
&= \frac{\sigma_d^2 \mathbb{E}[\|\mathbf{D}\|_F^2] + \sigma_p^2 \|\mathbf{P}\|_F^2}{\mathbb{E}[\|\mathbf{N}\|_F^2]} \\
&= \frac{MN \sigma_d^2 + \sigma_p^2}{MN \sigma^2} = \frac{\sigma_d^2 \left(1 + \frac{\beta}{MN}\right)}{\sigma^2},
\end{split}
\end{equation}
where $\beta = \sigma_p^2 / \sigma_d^2$ is the \gls{PDR} and allows for boosting the pilot amplitude to facilitate channel estimation \cite{Bello2025}.

\subsection{SP Channel Estimation Baseline: Threshold Method}
As a baseline for channel estimation in \gls{OFDM} systems using delay-Doppler superimposed pilots, the method proposed in \cite{Bello2025} is considered. This method, referred to as the \gls{TM}, relies on two simplifying assumptions: \gls{ICI}-free operation and integer delays and Doppler shifts.

The received delay-Doppler samples $\mathbf{Y}_{dd}$ are modeled as a circular convolution \cite{Bello2025,Viterbo2022}:
\begin{equation}\label{eq:2DcircconvDD}
\mathbf{Y}_{dd} = \mathbf{F}_M^\mathsf{H} \mathbf{Y} \mathbf{F}_N = \mathbf{H}_{dd} \circledast \mathbf{X}_{dd} + \mathbf{N}_{dd},
\end{equation}
where $\mathbf{H}_{dd} = \mathbf{F}_M^\mathsf{H} \mathbf{H}_{\mathrm{tf}} \mathbf{F}_N$ is the delay-Doppler channel matrix, and $\mathbf{X}_{dd} = \sigma_p \mathbf{P} + \sigma_d \mathbf{F}_M^\mathsf{H} \mathbf{D} \mathbf{F}_N$ are the delay-Doppler samples.

Assuming integer delays and Doppler shifts \cite{Bello2025}, such that $\tau_p = l_p / (M \Delta f)$ and $\nu_p = k_p / (N T')$, where $l_p, k_p \in \mathbb{Z}$ denote the normalized delay and Doppler indices, the relation in \eqref{eq:2DcircconvDD} simplifies to
\begin{equation}\label{eq:inoutrelDDintDD}
[\mathbf{Y}_{dd}]_{m,n} = \sum_{p=1}^P \alpha_p [\mathbf{X}_{dd}]_{[m-l_p]_M,[n-k_p]_N} + [\mathbf{N}_{dd}]_{m,n}.
\end{equation}
Using the delay-Doppler superimposed pilot, channel parameters $\{\alpha_p, l_p, k_p\}_{p=1}^P$ are estimated by detecting peaks that exceed a predefined threshold $\mathcal{T}$ \cite{Bello2025,kanazawa2025}. Specifically, a path with delay $\hat{l}_p$ and Doppler $\hat{k}_p$ is detected if
\begin{equation}
[\mathbf{Y}_{dd}]_{m_p+\hat{l}_p,n_p+\hat{k}_p} \geq \mathcal{T},
\end{equation}
and the channel gain is estimated from \eqref{eq:inoutrelDDintDD} as
\begin{equation}
\hat{\alpha}_p = \frac{[\mathbf{Y}_{dd}]_{m_p+\hat{l}_p,n_p+\hat{k}_p}}{\sigma_p}.
\end{equation}
The time-frequency channel matrix is then reconstructed as
\begin{equation}
\hat{\mathbf{H}}_{\mathrm{tf}} = \sum_{p=1}^{\hat{P}} \hat{\alpha}_p \mathbf{b}\Big(\frac{\hat{l}_p}{M \Delta f}\Big) \mathbf{c}^\top\Big(\frac{\hat{k}_p}{N T'}\Big),
\end{equation}
where $\hat{P}$ is the number of detected peaks exceeding the threshold $\mathcal{T}$. Following \cite{kanazawa2025}, the threshold is typically set as $\mathcal{T} = 3 \sqrt{\sigma^2 + \sigma_d^2}$.

\begin{remark}
When channel parameters are not integers, the pilot signal spreads across adjacent delay-Doppler bins \cite{Viterbo2022,Khan2021,Khan2023,marchese2024,Marchese2025disjoint}. This is the main limitation of the threshold method: a single multipath component with fractional parameters is approximated as multiple integer-valued components. Since only portions of the spread pilot exceeding the threshold are detected, some fractional multipath contributions may be missed. In contrast, channel estimation methods that explicitly account for fractional delays and Doppler shifts can accurately reconstruct the spreading effect, significantly improving estimation performance.
\end{remark}

\subsection{Proposed SP Channel Estimation Method Accounting for Fractional Channel Parameters}
This section introduces the proposed \gls{CE} scheme based on the \gls{ICI}-aware model in \eqref{eq:delaytimeObservations}.

\subsubsection{Single-path Scenario}
Considering a single-path channel and given the pilot model in \eqref{eq:PilotMtx}, disjoint delay-Doppler estimation can be performed as follows. The received delay-Doppler matrix is given by
\begin{equation}\label{eq:DDmtxSingleP}
\mathbf{Y}_{dd} = \mathbf{R} \mathbf{F}_N = \alpha \mathbf{\tilde{C}}(\nu) \mathbf{F}_{M}^{\mathsf{H}} \Big( \mathbf{X} \odot \mathbf{b}(\tau)\mathbf{c}^\top(\nu) \Big) \mathbf{F}_N + \mathbf{N}_{dd}.
\end{equation}
The delay and Doppler of the path are first initialized by searching for the maximum pilot energy in the delay-Doppler matrix \cite{Marchese2025disjoint,Khan2021,Khan2023}:
\begin{equation}\label{eq:DDinit}
\hat{l}, \hat{k} = \arg\max_{l,k} \Big| [\mathbf{Y}_{dd}]_{m_p + l, n_p + k} \Big|^2.
\end{equation}
The delay and Doppler can then be separately refined using the delay and Doppler profiles. The delay profile is obtained from the $k$-th column of the received delay-Doppler matrix \cite{Khan2021,Khan2023}:
\begin{equation}\label{eq:delayprofile}
\mathbf{u} = [\mathbf{Y}_{dd}]_{:, n_p+\hat{k}} = \mathbf{Y}_{dd} \mathbf{e}_{n_p+\hat{k}} \in \mathbb{C}^{M}.
\end{equation}
Similarly, the Doppler profile is obtained from the $l$-th row of the received delay-Doppler matrix:
\begin{equation}\label{eq:dopplerprofile}
\mathbf{v} = [\mathbf{Y}_{dd}]_{m_p\hat{l}, :} = \mathbf{Y}_{dd}^\top \mathbf{e}_{m_p+\hat{l}} \in \mathbb{C}^{N}.
\end{equation}
The delay-Doppler matrix can be decomposed as
\begin{equation}
\mathbf{Y}_{dd} = \mathbf{Y}_{dd}^{\text{pilot}} + \mathbf{Y}_{dd}^{\text{data}} + \mathbf{N}_{dd},
\end{equation}
where
\begin{equation}
\mathbf{Y}_{dd}^{\text{pilot}} = \alpha \sigma_p \mathbf{\tilde{C}}(\nu) \mathbf{F}_M^{\mathsf{H}} \Big[ \mathbf{F}_M \mathbf{P} \mathbf{F}_N^{\mathsf{H}} \odot \mathbf{b}(\tau) \mathbf{c}^\top(\nu) \Big] \mathbf{F}_N,
\end{equation}
\begin{equation}
\mathbf{Y}_{dd}^{\text{data}} = \alpha \sigma_d \mathbf{\tilde{C}}(\nu) \mathbf{F}_M^{\mathsf{H}} \Big[ \mathbf{D} \odot \mathbf{b}(\tau) \mathbf{c}^\top(\nu) \Big] \mathbf{F}_N.
\end{equation}
The delay profile can then be expressed as
\begin{equation}
\mathbf{u} = (\mathbf{Y}_{dd}^{\text{pilot}} + \mathbf{Y}_{dd}^{\text{data}}) \mathbf{e}_{n_p+\hat{k}} + \mathbf{n}_u,
\end{equation}
where $\mathbf{n}_u \sim \mathcal{CN}(\mathbf{0}, \sigma^2 \mathbf{I}_M)$. Specifically,
\begin{equation}\begin{split} \mathbf{Y}_{dd}&^{\text{pilot}}\mathbf{e}_{n_p+\hat{k}} \\ =&\alpha\sigma_p\mathbf{\tilde{C}}(\nu)\mathbf{F}_M^{\mathsf{H}}\Big[\mathbf{F}_M\mathbf{P}\mathbf{F}_N^{\mathsf{H}}\odot\mathbf{b}(\tau)\mathbf{c}^\top(\nu)\Big]\underbrace{\mathbf{F}_N\mathbf{e}_{n_p+\hat{k}}}_{\mathbf{f}_{n_p+\hat{k}}} 
\\ =&\alpha\sigma_p\mathbf{\tilde{C}}(\nu)\mathbf{F}_M^{\mathsf{H}}\Big[\underbrace{\mathbf{F}_M\mathbf{e}_{m_p}}_{\mathbf{f}_{m_p}}\underbrace{\mathbf{e}_{n_p}^\top\mathbf{F}_N^{\mathsf{H}}}_{(\mathbf{f}_{n_p}^*)^\top}\odot\mathbf{b}(\tau)\mathbf{c}^\top(\nu)\Big]\mathbf{f}_{n_p+\hat{k}} \\ =&\alpha\sigma_p\mathbf{\tilde{C}}(\nu)\mathbf{F}_M^{\mathsf{H}}\Big[\mathbf{f}_{m_p}\odot\mathbf{b}(\tau)\Big]\underbrace{\Big[\mathbf{f}_{n_p}^*\odot\mathbf{c}(\nu)\Big]^\top\mathbf{f}_{n_p+\hat{k}}}_{R_u} \\ =&\alpha\sigma_pR_u\mathbf{\tilde{C}}(\nu)\underbrace{\mathbf{F}_M^{\mathsf{H}}\big(\mathbf{f}_{m_p}\odot\mathbf{b}(\tau)\big)}_{\text{delay term}}. \end{split} \end{equation}
The correlation term $R_u$ represents a Doppler filtering process that extracts delay signals in the $\hat{k}$-th Doppler bin. The delay can then be estimated by correlating the delay profile with the delay term \cite{Khan2021,Khan2023}:
\begin{equation}\label{eq:delayxcorr}
\hat{\tau} = \arg\max_\tau \Big| \mathbf{u}^\mathsf{H} \mathbf{F}_M^{\mathsf{H}} \big( \mathbf{f}_{m_p} \odot \mathbf{b}(\tau) \big) \Big|.
\end{equation}
This procedure allows refining the delay estimate in a neighborhood around $\hat{l} / (M \Delta f)$.

Similarly, the Doppler profile is obtained as
\begin{equation}
\mathbf{v} = \big( {\mathbf{Y}_{dd}^{\text{pilot}}}^{\top} + {\mathbf{Y}_{dd}^{\text{data}}}^{\top} \big) \mathbf{e}_{m_p+\hat{l}} + \mathbf{n}_v,
\end{equation}
where $\mathbf{n}_v \sim \mathcal{CN}(\mathbf{0}, \sigma^2 \mathbf{I}_N)$. Specifically,
\begin{equation}\begin{split} {\mathbf{Y}_{dd}^{\text{pilot}}}&^{\top}\mathbf{e}_{m_p+\hat{l}} \\ =&\alpha\sigma_p\mathbf{F}_N\Big[\mathbf{F}_M\mathbf{P}\mathbf{F}_N^{\mathsf{H}}\odot\mathbf{b}(\tau)\mathbf{c}^\top(\nu)\Big]^\top\underbrace{\mathbf{F}_M^{\mathsf{H}}\mathbf{\tilde{C}}(\nu)\mathbf{e}_{m_p+\hat{l}}}_{\approx \mathbf{F}_M^{\mathsf{H}}\mathbf{e}_{m_p+\hat{l}}=\mathbf{f}_{m_p+\hat{l}}} \\ \approx&\alpha\sigma_p\mathbf{F}_N\Big[\mathbf{F}_N^{\mathsf{H}}\mathbf{P}^\top\mathbf{F}_M\odot\mathbf{c}(\nu)\mathbf{b}^\top(\tau)\Big]\mathbf{f}_{m_p+\hat{l}} \\ =&\alpha\sigma_p\mathbf{F}_N\Big[\mathbf{F}_N^{\mathsf{H}}\mathbf{e}_{n_p}\mathbf{e}_{m_p}^\top\mathbf{F}_M\odot\mathbf{c}(\nu)\mathbf{b}^\top(\tau)\Big]\mathbf{f}_{m_p+\hat{l}} \\ =&\alpha\sigma_p\mathbf{F}_N\Big[\mathbf{f}^*_{n_p}\odot\mathbf{c}(\nu)\Big]\underbrace{\Big[\mathbf{f}_{m_p}\odot\mathbf{b}(\tau)\Big]^\top\mathbf{f}_{m_p+\hat{l}}}_{R_v} \\ =&\alpha\sigma_pR_v\underbrace{\mathbf{F}_N\big(\mathbf{f}^*_{n_p}\odot\mathbf{c}(\nu)\big)}_{\text{Doppler term}}. \end{split} \end{equation}
The correlation term $R_v$ represents a delay filtering process that extracts Doppler signals in the $\hat{l}$-th delay bin. Similarly to the delay case, the Doppler can be refined by correlating the Doppler profile with the Doppler term \cite{Khan2021,Khan2023}:
\begin{equation}\label{eq:dopplerxcorr}
\hat{\nu} = \arg\max_\nu \Big| \mathbf{v}^\mathsf{H} \mathbf{F}_N \big( \mathbf{f}^*_{n_p} \odot \mathbf{c}(\nu) \big) \Big|.
\end{equation}
This procedure refines the Doppler estimate in a neighborhood around $\hat{k} / (N T')$. Once the delay and Doppler are estimated, the optimal channel gain is obtained via \gls{LS} as
\begin{equation}\label{eq:LSformulation}
\hat{\alpha} = \arg\min_\alpha \Big\| \mathbf{R} - \alpha \mathbf{A}(\hat\tau, \hat\nu) \Big\|_F^2,
\end{equation}
where
\begin{equation}
\mathbf{A}(\tau, \nu) = \sigma_p \mathbf{\tilde{C}}(\nu) \mathbf{F}_{M}^{\mathsf{H}} \Big( \mathbf{F}_M \mathbf{P} \mathbf{F}_N^\mathsf{H} \odot \mathbf{b}(\tau) \mathbf{c}^\top(\nu) \Big).
\end{equation}
Solving \eqref{eq:LSformulation}, the optimal gain is
\begin{equation}\label{eq:chgainLSestimation}
\hat{\alpha} = \frac{\mathrm{Tr} \big( \mathbf{A}^{\mathsf{H}}(\hat{\tau}, \hat{\nu}) \mathbf{R} \big)}{\big\| \mathbf{A}(\hat{\tau}, \hat{\nu}) \big\|_F^2}.
\end{equation}

\subsubsection{Multipath Scenario}
In the case of multiple paths, the channel parameters of each path are estimated sequentially \cite{marchese2024,Marchese2025disjoint,Khan2021,Khan2023} using the residual delay-Doppler matrix $\boldsymbol{\mathcal{E}}_{dd}^{(p)}$. In the first iteration, $\boldsymbol{\mathcal{E}}_{dd}^{(0)} = \mathbf{Y}_{dd}$. During the $p$-th iteration, the delay and Doppler of the $p$-th path are initialized by searching for the maximum pilot energy in the residual delay-Doppler matrix $\boldsymbol{\mathcal{E}}_{dd}^{(p-1)}$ as in \eqref{eq:DDinit}. Subsequently, the delay and Doppler are separately refined using \eqref{eq:delayxcorr} and \eqref{eq:dopplerxcorr} within a neighborhood of $\hat{l}_p / M \Delta f$ and $\hat{k}_p / N T'$, respectively. The channel gain is then estimated from \eqref{eq:chgainLSestimation} as
\begin{equation}\label{eq:chgainLSestimation_multipath}
\hat{\alpha}_p = \frac{\mathrm{Tr} \big( \mathbf{A}^{\mathsf{H}}(\hat{\tau}_p, \hat{\nu}_p) \boldsymbol{\mathcal{E}}_{dd}^{(p-1)} \mathbf{F}_N^{\mathsf{H}} \big)}{\big\| \mathbf{A}(\hat{\tau}_p, \hat{\nu}_p) \big\|_F^2}.
\end{equation}
Once the parameters of a path are obtained, the next path is estimated on the updated residual matrix
\begin{equation}
\boldsymbol{\mathcal{E}}_{dd}^{(p)} = \boldsymbol{\mathcal{E}}_{dd}^{(p-1)} - \hat{\alpha}_p \mathbf{A}(\hat{\tau}_p, \hat{\nu}_p) \mathbf{F}_N.
\end{equation}
This subtraction removes the contribution of the current path and mitigates \gls{IPI} arising from fractional effects \cite{marchese2024,Marchese2025disjoint,Khan2021,Khan2023}. The iterative procedure continues until either a maximum number of paths $P_{\max}$ is estimated or the stopping criterion
\begin{equation}
\big\| \boldsymbol{\mathcal{E}}_{dd}^{(p)} - \boldsymbol{\mathcal{E}}_{dd}^{(p-1)} \big\|_F < \epsilon,
\end{equation}
for a small arbitrary $\epsilon$, is satisfied. The total number of detected paths is then $\hat{P} = \min(p, P_{\max})$.

The proposed multipath estimation approach is summarized in Algorithm \ref{alg:propCEalg}.

\begin{algorithm}[t]
\caption{Proposed Fractional \gls{CE} Algorithm}\label{alg:propCEalg}
\KwIn{$\mathbf{Y}_{dd},\ \mathbf{P},\ \ P_{\max},\ \epsilon$}
$\boldsymbol{\mathcal{E}}_{dd}^{(0)}=\mathbf{Y}_{dd}$\;
\For{$p = 1$ \KwTo $P_{\max}$}{
$\hat{l}_p,\hat{k}_p=\arg\max_{l,k} \Big|\big[\boldsymbol{\mathcal{E}}_{dd}^{(p-1)}\big]_{m_p+l,n_p+k}\Big|^2$\;

$\mathbf{u}=\Big[\boldsymbol{\mathcal{E}}_{dd}^{(p-1)}\Big]_{:,n_p+\hat{k}_p}$\;
$\hat{\tau}_p=\arg\max_\tau\Big|\mathbf{u}^\mathsf{H}\mathbf{F}_M^{\mathsf{H}}\big(\mathbf{f}_{m_p}\odot\mathbf{b}(\tau)\big)\Big|$\;

$\mathbf{v}=\Big[\boldsymbol{\mathcal{E}}_{dd}^{(p-1)}\Big]_{m_p+\hat{l}_p,:}$\;
$\hat{\nu}_p=\arg\max_\nu\Big|\mathbf{v}^\mathsf{H}\mathbf{F}_N\big(\mathbf{f}^*_{n_p}\odot\mathbf{c}(\nu)\big)\Big|$\;

$\hat{\alpha}_p = \frac{\mathrm{Tr}\big(\mathbf{A}^{\mathsf{H}}(\hat\tau_p,\hat\nu_p)\boldsymbol{\mathcal{E}}_{dd}^{(p-1)}\mathbf{F}_N^{\mathsf{H}}\big)}{\big\|\mathbf{A}(\hat\tau_p,\hat\nu_p)\big\|_F^2}$\;

$\boldsymbol{\mathcal{E}}_{dd}^{(p)}=\boldsymbol{\mathcal{E}}_{dd}^{(p-1)}-\hat\alpha_p\mathbf{A}(\hat\tau_p,\hat\nu_p)\mathbf{F}_N$\;

    \If{$\big\|\boldsymbol{\mathcal{E}}_{dd}^{(p)}-\boldsymbol{\mathcal{E}}_{dd}^{(p-1)}\big\|_F<\epsilon$}{
            $\hat{P} = p$\;
            \textbf{break}\;
    }
}
\textbf{Output:}~$\{\hat{\alpha}_{p},\hat{\tau}_p,\hat{\nu}_p\}_{p=1}^{\hat{P}}$\;
\end{algorithm}

\section{Symbol Detection}
Assuming knowledge of the \gls{CSI}, this section reviews conventional detection schemes and introduces the proposed iterative method for symbol detection. Specifically, the goal of symbol detection is to estimate the time-frequency symbol matrix $\mathbf{X}$. In the case of \gls{SP}, after estimating the time-frequency symbol matrix according to the model in \eqref{eq:SuperimposedPilotModel}, an estimate of the data symbols is obtained by removing the pilot contribution as
\begin{equation}
\hat{\mathbf{D}} = \frac{\hat{\mathbf{X}} - \sigma_p \mathbf{F}_M \mathbf{P} \mathbf{F}_N^\mathsf{H}}{\sigma_d}.
\end{equation}

\subsection{Baseline Methods}
\subsubsection{Single-Tap MMSE (\gls{ICI}-free)}
Under the assumption of negligible \gls{ICI}, the model in \eqref{eq:ICIfreeModel} applies. The receiver can perform single-tap \gls{MMSE} equalization to estimate $\mathbf{X}$ as
\begin{equation}
\hat{\mathbf{X}} = \frac{\mathbf{H}_{tf}^* \odot \mathbf{Y}}{|\mathbf{H}_{tf}|^2 + \text{SNR}^{-1}}.
\end{equation}
The overall complexity of this approach for the entire \gls{OFDM} frame is $\mathcal{O}(MN)$.

\subsubsection{Full MMSE (\gls{ICI}-aware)}
For non-negligible \gls{ICI}, the receiver must perform full \gls{MMSE} equalization \cite{Choi2001,Yiyan2025} to compensate for the loss of orthogonality between subcarriers
\begin{equation}
\hat{\mathbf{x}}_n = \Big(\mathbf{H}_{tf,n}^\mathsf{H} \mathbf{H}_{tf,n} + \text{SNR}^{-1} \mathbf{I}_M \Big)^{-1} \mathbf{H}_{tf,n}^\mathsf{H} \mathbf{y}_n,
\end{equation}
where $\mathbf{y}_n = \mathbf{F}_M \mathbf{r}_n$ and $\mathbf{H}_{tf,n} = \mathbf{F}_M \mathbf{H}_n \mathbf{F}_M^\mathsf{H}$. The estimates are then concatenated as $\hat{\mathbf{X}} = [\hat{\mathbf{x}}_0 \ \hat{\mathbf{x}}_1 \ \dots \ \hat{\mathbf{x}}_{N-1}]$.

The overall complexity of full \gls{MMSE} detection for the entire \gls{OFDM} frame is $\mathcal{O}(NM^3)$ due to the matrix inversion.

\begin{algorithm}[t]
\caption{Proposed IMFC Equalizer based on Landweber Method}\label{alg:LandweberDet}
\KwIn{$\mathbf{R},\ \Big\{{\mathbf{H}}_{tf}^{(p)}, {\mathbf{H}}_{\text{ICI}}^{(p)}, {\alpha}_p\Big\}_{p=1}^P,\ \eta$}
$\hat{\mathbf{X}}^{(0)}=\mathbf{0}$\;
$\mathbf{E}^{(0)}=\mathbf{R}$\;
\For{$t = 1$ \KwTo $T$}{

    $\Delta\hat{\mathbf{X}}^{(t)} = \mathbf{0}$\;

    \For{$p = 1$ \KwTo $P$}{
        $\hat{\mathbf{X}}_p^{(t)} \leftarrow \mathbf{E}^{(t-1)}\odot\Big({\mathbf{H}}_{\text{ICI}}^{(p)}\Big)^*$\;

        $\hat{\mathbf{X}}_p^{(t)} \leftarrow \mathbf{F}_M\hat{\mathbf{X}}_p^{(t)}$\;

        $\hat{\mathbf{X}}_p^{(t)} \leftarrow \hat{\mathbf{X}}_p^{(t)}\odot\Big({\mathbf{H}}_{tf}^{(p)}\Big)^*$\;

        $\Delta\hat{\mathbf{X}}^{(t)} \leftarrow \Delta\hat{\mathbf{X}}^{(t)} + {\alpha}_p^*\hat{\mathbf{X}}_p^{(t)}$\;
    }

$\hat{\mathbf{X}}^{(t)}=\hat{\mathbf{X}}^{(t-1)}+\eta\Delta\hat{\mathbf{X}}^{(t)}$\;

$\mathbf{E}^{(t)}=\mathbf{R}$\;

\For{$p = 1$ \KwTo $P$}{
 $\mathbf{E}_p^{(t)}\leftarrow \hat{\mathbf{X}}^{(t)}\odot \mathbf{H}_{tf}^{(p)}$

    $\mathbf{E}_p^{(t)}\leftarrow \mathbf{F}_{M}^{\mathsf{H}}\mathbf{E}_p^{(t)}$\;

    $\mathbf{E}_p^{(t)} \leftarrow \mathbf{E}_p^{(t)}\odot{\mathbf{H}}_{\text{ICI}}^{(p)}$\;
    
    $\mathbf{E}^{(t)} \leftarrow \mathbf{E}^{(t)} - {\alpha}_p \cdot \mathbf{E}_p^{(t)}$\;
} 

}
\textbf{Output:}~$\hat{\mathbf{X}}^{(\text{T})}$\;
\end{algorithm}

\subsection{Proposed Reduced-Complexity Least Squares Detection via Landweber Method}
This section introduces the proposed low-complexity equalizer based on the Landweber method, referred to as \gls{IMFC}. The \gls{LS} detection problem is formulated as
\begin{equation}\label{eq:detProbFormulation}
\hat{\mathbf{X}} = \arg\min_{\mathbf{X}} \Big\| \mathbf{R} - \boldsymbol{\mathcal{H}}(\mathbf{X}) \Big\|_F^2. 
\end{equation}
Given the problem in \eqref{eq:detProbFormulation}, the Landweber iterative update is
\begin{equation}\label{eq:LandweberMethod}
\hat{\mathbf{X}}^{(t)} = \hat{\mathbf{X}}^{(t-1)} + \eta \boldsymbol{\mathcal{H}}^\mathsf{H} \Big( \mathbf{R} - \boldsymbol{\mathcal{H}}(\hat{\mathbf{X}}^{(t-1)}) \Big),
\end{equation}
where $\eta$ is the step-size parameter and the initial estimate is $\hat{\mathbf{X}}^{(0)} = \mathbf{0}$. It can be noted that the learning rate $\eta$ must be properly selected to guarantee convergence of the Landweber iteration \cite{CharlesByrne_2004}. In practice, $\eta$ can be chosen empirically or adapted through a decay model, in accordance with standard results on iterative reconstruction methods.
Exploiting the structure of the channel operator $\boldsymbol{\mathcal{H}}(\cdot)$, the method can be decomposed into low-complexity path-wise operations. During the $t$-th iteration, the algorithm performs the following steps:

\subsubsection{Single-Tap \gls{ICI} Compensation in Delay-Time Domain}
\begin{equation}
\hat{\mathbf{X}}_p^{(t)} \leftarrow \mathbf{E}^{(t-1)} \odot \big(\mathbf{H}_{\text{ICI}}^{(p)}\big)^*.
\end{equation}

\subsubsection{Conversion to Frequency-Time Domain}
\begin{equation}
\hat{\mathbf{X}}_p^{(t)} \leftarrow \mathbf{F}_M \hat{\mathbf{X}}_p^{(t)}.
\end{equation}

\subsubsection{Single-Tap Frequency-Time Channel Equalization}
\begin{equation}
\hat{\mathbf{X}}_p^{(t)} \leftarrow \hat{\mathbf{X}}_p^{(t)} \odot \big(\mathbf{H}_{tf}^{(p)}\big)^*.
\end{equation}

\subsubsection{Combining}
The $P$ path-wise estimates $\hat{\mathbf{X}}_p^{(t)}$ are combined using \gls{MRC} to maximize the \gls{SNR}:
\begin{equation}
\Delta \hat{\mathbf{X}}^{(t)} = \sum_{p=1}^P \alpha_p^* \hat{\mathbf{X}}_p^{(t)}.
\end{equation}
The Landweber update is then applied as
\begin{equation}
\hat{\mathbf{X}}^{(t)} = \hat{\mathbf{X}}^{(t-1)} + \eta \Delta \hat{\mathbf{X}}^{(t)},
\end{equation}
and the residual observation matrix is updated as
\begin{equation}
\mathbf{E}^{(t)} = \mathbf{R} - \boldsymbol{\mathcal{H}}(\hat{\mathbf{X}}^{(t)}),
\end{equation}
and it can ben noted that $\mathbf{E}^{(0)} = \mathbf{R}$ since the symbol matrix is initialized as zero.

The pseudocode of the proposed \gls{IMFC} method is provided in Algorithm \ref{alg:LandweberDet}. The overall complexity of this iterative procedure, assuming $\log_2 M < N$, is dominated by the single-tap operations and scales as $\mathcal{O}(TPMN)$, where $T$ is the number of iterations.

\begin{table}[t]
\centering
\caption{Simulation parameters.}\label{TabSim}
{
\begin{tabular}{|l | l|}
\hline
\multicolumn{2}{|c|}{\textbf{General}} \\
\hline
Carrier frequency, $f_c$ & $5.9$ GHz \\
\hline
Number of subcarriers, $M$ & $128$ \\
\hline
Number of \gls{OFDM} symbols, $N$ & $32$ \\
\hline
Subcarrier spacing, $\Delta f$ & $30$ kHz \\
\hline
\gls{CP} duration, $T_{\text{CP}}$ & $5 \ \mu$s \\
\hline
Modulation & \gls{QAM}, $Q=4,16$ \\
\hline
\multicolumn{2}{|c|}{\textbf{Wireless channel}} \\
\hline
Number of multipaths, $P$ & $4$ \\
\hline
Propagation delays, $\tau_p$ & $[0, 0.9, 2.7, 4] \ \mu$s \\
\hline
Power delay profile & Uniform \\
\hline
Doppler shifts, $\nu_p$ & \makecell[l]{$\nu_p=f_c\frac{v_{\max}}{c}\cos(\theta)$ \\ $\theta\sim\mathcal{U}[0,2\pi]$} \\
\hline
\end{tabular}
}
\end{table}

\section{Simulation Results}
This section presents the simulation results used to validate the proposed receiver architecture. The main simulation parameters are listed in Table \ref{TabSim}. Two different channel models are considered:

\begin{itemize}
    \item Fractional channel: This channel model resembles a realistic environment with fractional delays and Doppler shifts. Channel parameters are as listed in Table \ref{TabSim}.
    \item Integer channel: This channel assumes integer delays and Doppler shifts and is used to compare the proposed scheme against the baseline method in \cite{Bello2025}, which relies on this assumption. The channel parameters are obtained from Table \ref{TabSim} as
    \begin{equation}
        \tau_p^{\text{int}} = \frac{\text{round}(\tau_p M \Delta f)}{M \Delta f}, \quad
        \nu_p^{\text{int}} = \frac{\text{round}(\nu_p N T')}{N T'}.
    \end{equation}
\end{itemize}

The following receiver architectures are considered for numerical evaluation:

\begin{itemize}
    \item \textbf{TM + Full MMSE:} Channel estimation via the \gls{TM} method \cite{Bello2025} using the delay-Doppler \gls{SP} scheme and equalization via the \gls{ICI}-aware full \gls{MMSE} equalizer.
    
    \item \textbf{Prop. CE + Full MMSE:} Channel estimation via the proposed fractional estimation method (Algorithm \ref{alg:propCEalg}) with the delay-Doppler \gls{SP} scheme, equalization via the \gls{ICI}-aware full \gls{MMSE} equalizer.
    
    \item \textbf{Perf. CSI + Full MMSE:} Delay-Doppler \gls{SP} scheme is adopted at the transmitter, perfect \gls{CSI} is assumed at the receiver, and equalization is performed using the \gls{ICI}-aware full \gls{MMSE} equalizer. This serves as an upper bound on communication performance.
    
    \item \textbf{TM + Single-Tap:} Channel estimation via the \gls{TM} method with the delay-Doppler \gls{SP} scheme and \gls{ICI}-free single-tap \gls{MMSE} equalization. This corresponds to the baseline receiver in \cite{Bello2025}.
    
    \item \textbf{EP + Single-Tap:} Channel estimation via conventional \gls{EP} in \gls{OFDM} systems and equalization via the \gls{ICI}-free single-tap \gls{MMSE} equalizer.
    
    \item \textbf{Prop. CE + Prop. IMFC:} Channel estimation via the proposed fractional estimation method (Algorithm \ref{alg:propCEalg}) with the delay-Doppler \gls{SP} scheme and equalization via the proposed \gls{ICI}-aware \gls{IMFC} equalizer.
\end{itemize}

A complexity comparison of all schemes is provided in Table \ref{TabComplexityMethods}. It can be observed that the TM + Single-Tap and conventional EP + Single-Tap schemes have the lowest complexity, at the cost of neglecting \gls{ICI} and fractional effects. In contrast, the Prop. CE + Prop. IMFC scheme has higher complexity but remains linear in the number of \glspl{RE}, unlike full \gls{MMSE}-based receivers, whose equalization complexity scales cubically with the number of \gls{OFDM} subcarriers.

\begin{table}[t]
\centering
\caption{Complexity comparison of the different methods.}\label{TabComplexityMethods}
{
\begin{tabular}{|l|c|c|}
\hline
\textbf{Method} & \textbf{CE} & \textbf{Equalization} \\
\hline
TM + Full MMSE & $\mathcal{O}(MN)$ & $\mathcal{O}(NM^3)$ \\
\hline
Prop. CE + Full MMSE & $\mathcal{O}(P_{\max}MN)$ & $\mathcal{O}(NM^3)$ \\
\hline
Perf. CSI + Full MMSE & - & $\mathcal{O}(NM^3)$ \\
\hline
TM + Single-Tap & $\mathcal{O}(MN)$ & $\mathcal{O}(MN)$ \\
\hline
EP + Single-Tap & $\mathcal{O}(MN)$ & $\mathcal{O}(MN)$ \\
\hline
Prop. CE + Prop. IMFC & $\mathcal{O}(P_{\max}MN)$ & $\mathcal{O}(TPMN)$ \\
\hline
\end{tabular}
}
\end{table}

\subsection{Peak-to-Average Power Ratio}
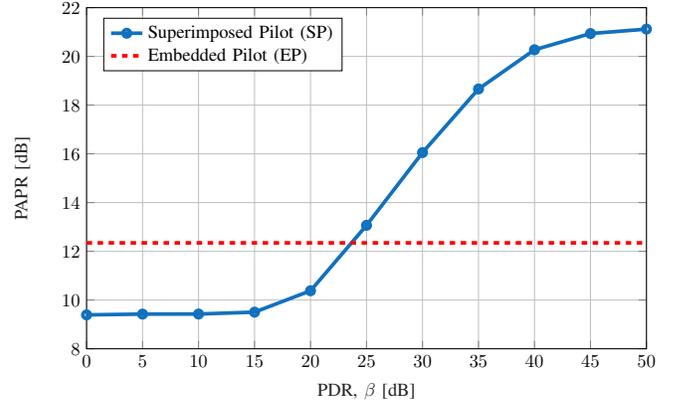
\begin{figure}[t]
 \centering
 \resizebox{0.99\columnwidth}{!}{
%
%
\definecolor{mycolor1}{rgb}{0.06600,0.44300,0.74500}%
\definecolor{mycolor2}{rgb}{0.12941,0.12941,0.12941}%
\begin{tikzpicture}

\begin{axis}[%
width=4.225in,
height=2.576in,
scale only axis,
xmin=0,
xmax=50,
xlabel style={font=\color{mycolor2}},
xlabel={PDR, $\beta$ [dB]},
ymin=8,
ymax=22,
ylabel style={font=\color{mycolor2}},
ylabel={PAPR [dB]},
axis background/.style={fill=white},
xmajorgrids,
ymajorgrids,
legend style={at={(0.03,0.97)}, anchor=north west, legend cell align=left, align=left}
]
\addplot [color=mycolor1, line width=2.0pt, mark=o, mark options={solid, mycolor1}]
  table[row sep=crcr]{%
0	9.39097442264887\\
5	9.42292575512315\\
10	9.42576527449495\\
15	9.50351124576593\\
20	10.3812133490604\\
25	13.0666172885656\\
30	16.0519833351014\\
35	18.6630578359195\\
40	20.271814223197\\
45	20.9379198869566\\
50	21.1201423829979\\
};
\addlegendentry{Superimposed Pilot (SP)}

\addplot [color=red, dashed, line width=2.0pt]
  table[row sep=crcr]{%
0	12.3466210668473\\
50	12.3466210668473\\
};
\addlegendentry{Embedded Pilot (EP)}

\end{axis}

\end{tikzpicture}%
}
 \caption{The PAPR of the delay-Doppler \gls{SP}-based scheme is shown for different values of \gls{PDR}and compared against the PAPR of the \gls{EP}-based scheme} \label{fig:PAPR}
\end{figure}

The \gls{PAPR} is a critical metric in modern wireless systems, particularly for multi-carrier schemes. High \gls{PAPR} increases the linearity requirements of the \gls{PA}, forcing it to operate with a large power back-off. This reduces efficiency, increases energy consumption, and may introduce non-linear distortion that degrades signal reliability and spectral purity. Therefore, evaluating the \gls{PAPR} of different transmission schemes is essential for assessing their practical feasibility in real-world deployments.
The \gls{PAPR} is defined as
\begin{equation}
\text{PAPR} = \frac{\max\limits_{m,n} \big|[\mathbf{S}]_{m,n}\big|^2}{P_T}.
\end{equation}
Figure \ref{fig:PAPR} reports the \gls{PAPR} of the delay-Doppler \gls{SP} scheme for different values of the \gls{PDR}. The \gls{PAPR} of the conventional \gls{EP} scheme is also shown for comparison. It can be observed that for \gls{PDR} values below $23$ dB, the \gls{SP} scheme exhibits a lower \gls{PAPR} than the \gls{EP} scheme, with a minimum \gls{PAPR} slightly below $10$ dB. Conversely, for higher \gls{PDR} values, the \gls{PAPR} increases and can reach up to $21$ dB.

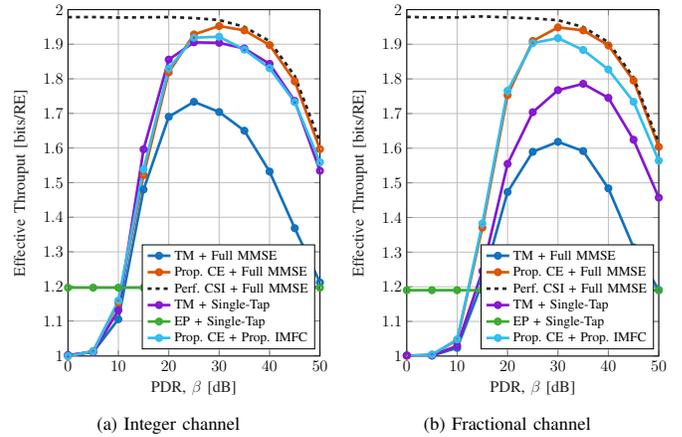
\begin{figure}[t]
 \centering
 \begin{subfigure}[t]{0.24\textwidth} 
 \centering
 \resizebox{\columnwidth}{!}{
%
%
\definecolor{mycolor1}{rgb}{0.06600,0.44300,0.74500}%
\definecolor{mycolor2}{rgb}{0.86600,0.32900,0.00000}%
\definecolor{mycolor3}{rgb}{0.12941,0.12941,0.12941}%
\definecolor{mycolor4}{rgb}{0.52100,0.08600,0.81900}%
\definecolor{mycolor5}{rgb}{0.23100,0.66600,0.19600}%
\definecolor{mycolor6}{rgb}{0.18400,0.74500,0.93700}%
\begin{tikzpicture}

\begin{axis}[%
width=2.61in,
height=3.591in,
scale only axis,
xmin=0,
xmax=50,
xlabel style={font=\color{mycolor3}},
xlabel={\large PDR, $\beta$ [dB]},
ymin=0.998757302761078,
ymax=2,
ylabel style={font=\color{mycolor3}},
ylabel={\large Effective Throuput [bits/RE]},
axis background/.style={fill=white},
title style={font=\bfseries\color{mycolor3}},
ticklabel style={font=\large},
xmajorgrids,
ymajorgrids,
legend style={at={(0.98,0.02)}, anchor=south east, legend cell align=left, align=left}
]
\addplot [color=mycolor1, line width=2.0pt, mark=o, mark options={solid, mycolor1}]
  table[row sep=crcr]{%
0	1.00135009765625\\
5	1.01032958984375\\
10	1.10521240234375\\
15	1.4799462890625\\
20	1.69032470703125\\
25	1.7335498046875\\
30	1.704111328125\\
35	1.6500341796875\\
40	1.5318896484375\\
45	1.36829345703125\\
50	1.21136962890625\\
};
\addlegendentry{TM + Full MMSE}

\addplot [color=mycolor2, line width=2.0pt, mark=o, mark options={solid, mycolor2}]
  table[row sep=crcr]{%
0	0.99890625\\
5	1.01313232421875\\
10	1.1535791015625\\
15	1.52176025390625\\
20	1.818740234375\\
25	1.92824951171875\\
30	1.95244140625\\
35	1.93993408203125\\
40	1.8976318359375\\
45	1.79270263671875\\
50	1.59668701171875\\
};
\addlegendentry{Prop. CE + Full MMSE}

\addplot [color=mycolor3, dashed, line width=2.0pt]
  table[row sep=crcr]{%
0	1.97814453125\\
5	1.9783984375\\
10	1.976982421875\\
15	1.977666015625\\
20	1.978388671875\\
25	1.9754931640625\\
30	1.96987060546875\\
35	1.950263671875\\
40	1.908837890625\\
45	1.8083544921875\\
50	1.618173828125\\
};
\addlegendentry{Perf. CSI + Full MMSE}

\addplot [color=mycolor4, line width=2.0pt, mark=o, mark options={solid, mycolor4}]
  table[row sep=crcr]{%
0	1.00106689453125\\
5	1.0106591796875\\
10	1.13099853515625\\
15	1.5962451171875\\
20	1.85549072265625\\
25	1.90541015625\\
30	1.903935546875\\
35	1.88755859375\\
40	1.84329345703125\\
45	1.73600830078125\\
50	1.534208984375\\
};
\addlegendentry{TM + Single-Tap}

\addplot [color=mycolor5, line width=2.0pt, mark=o, mark options={solid, mycolor5}]
  table[row sep=crcr]{%
0	1.19640380859375\\
5	1.19640380859375\\
10	1.19640380859375\\
15	1.19640380859375\\
20	1.19640380859375\\
25	1.19640380859375\\
30	1.19640380859375\\
35	1.19640380859375\\
40	1.19640380859375\\
45	1.19640380859375\\
50	1.19640380859375\\
};
\addlegendentry{EP + Single-Tap}

\addplot [color=mycolor6, line width=2.0pt, mark=o, mark options={solid, mycolor6}]
  table[row sep=crcr]{%
0	0.99875732421875\\
5	1.01232421875\\
10	1.15905517578125\\
15	1.53833740234375\\
20	1.82961669921875\\
25	1.9188232421875\\
30	1.9215869140625\\
35	1.88427978515625\\
40	1.83072265625\\
45	1.73265380859375\\
50	1.55890380859375\\
};
\addlegendentry{Prop. CE + Prop. IMFC}

\end{axis}

\end{tikzpicture}%
}
 \caption{Integer channel}\label{fig:TvsPDR_int}
 \end{subfigure}%
 \hfill 
\begin{subfigure}[t]{0.24\textwidth} 
 \centering
 \resizebox{\columnwidth}{!}{
%
%
\definecolor{mycolor1}{rgb}{0.06600,0.44300,0.74500}%
\definecolor{mycolor2}{rgb}{0.86600,0.32900,0.00000}%
\definecolor{mycolor3}{rgb}{0.12941,0.12941,0.12941}%
\definecolor{mycolor4}{rgb}{0.52100,0.08600,0.81900}%
\definecolor{mycolor5}{rgb}{0.23100,0.66600,0.19600}%
\definecolor{mycolor6}{rgb}{0.18400,0.74500,0.93700}%
\begin{tikzpicture}

\begin{axis}[%
width=2.61in,
height=3.591in,
scale only axis,
xmin=0,
xmax=50,
xlabel style={font=\color{mycolor3}},
xlabel={\large PDR, $\beta$ [dB]},
ymin=1.00002443790436,
ymax=2,
ylabel style={font=\color{mycolor3}},
ylabel={\large Effective Throuput [bits/RE]},
axis background/.style={fill=white},
title style={font=\bfseries\color{mycolor3}},
ticklabel style={font=\large},
xmajorgrids,
ymajorgrids,
legend style={at={(0.98,0.02)}, anchor=south east, legend cell align=left, align=left}
]
\addplot [color=mycolor1, line width=2.0pt, mark=o, mark options={solid, mycolor1}]
  table[row sep=crcr]{%
0	1.00157470703125\\
5	1.001455078125\\
10	1.02348876953125\\
15	1.21293212890625\\
20	1.47346923828125\\
25	1.58953369140625\\
30	1.6183837890625\\
35	1.5917138671875\\
40	1.48409423828125\\
45	1.31353759765625\\
50	1.19058837890625\\
};
\addlegendentry{TM + Full MMSE}

\addplot [color=mycolor2, line width=2.0pt, mark=o, mark options={solid, mycolor2}]
  table[row sep=crcr]{%
0	1.00081787109375\\
5	1.0037841796875\\
10	1.04614013671875\\
15	1.37084716796875\\
20	1.75317626953125\\
25	1.90941650390625\\
30	1.94867919921875\\
35	1.9401611328125\\
40	1.8967041015625\\
45	1.79507080078125\\
50	1.60390380859375\\
};
\addlegendentry{Prop. CE + Full MMSE}

\addplot [color=mycolor3, dashed, line width=2.0pt]
  table[row sep=crcr]{%
0	1.97897705078125\\
5	1.97716796875\\
10	1.977587890625\\
15	1.98039306640625\\
20	1.97735107421875\\
25	1.9746826171875\\
30	1.96942138671875\\
35	1.94994873046875\\
40	1.9057861328125\\
45	1.804287109375\\
50	1.61546875\\
};
\addlegendentry{Perf. CSI + Full MMSE}

\addplot [color=mycolor4, line width=2.0pt, mark=o, mark options={solid, mycolor4}]
  table[row sep=crcr]{%
0	1.0017578125\\
5	1.001826171875\\
10	1.02962158203125\\
15	1.24516357421875\\
20	1.554697265625\\
25	1.7039501953125\\
30	1.767421875\\
35	1.78588623046875\\
40	1.7453369140625\\
45	1.6243798828125\\
50	1.45728515625\\
};
\addlegendentry{TM + Single-Tap}

\addplot [color=mycolor5, line width=2.0pt, mark=o, mark options={solid, mycolor5}]
  table[row sep=crcr]{%
0	1.19017223011364\\
5	1.19017223011364\\
10	1.19017223011364\\
15	1.19017223011364\\
20	1.19017223011364\\
25	1.19017223011364\\
30	1.19017223011364\\
35	1.19017223011364\\
40	1.19017223011364\\
45	1.19017223011364\\
50	1.19017223011364\\
};
\addlegendentry{EP + Single-Tap}

\addplot [color=mycolor6, line width=2.0pt, mark=o, mark options={solid, mycolor6}]
  table[row sep=crcr]{%
0	1.0000244140625\\
5	1.00421875\\
10	1.04808837890625\\
15	1.38261474609375\\
20	1.7659716796875\\
25	1.9032666015625\\
30	1.91771240234375\\
35	1.88358642578125\\
40	1.827109375\\
45	1.73391357421875\\
50	1.5639208984375\\
};
\addlegendentry{Prop. CE + Prop. IMFC}

\end{axis}

\end{tikzpicture}%
}
 \caption{Fractional channel}\label{fig:TvsPDR_frac}
 \end{subfigure}%

 \caption{Effective throughput against PDR for a fixed SNR of $15$ dB. The modulation order is $Q=4$ and the maximum speed is $1000$ km/h.} \label{fig:TvsPDR} 
\end{figure}

\subsection{Communication Performance Metrics}
To fairly compare different pilot schemes (\gls{SP} and \gls{EP}) that use time-frequency and power resources differently, the \emph{effective throughput} $\eta_{\text{eff}}$ is adopted as the main communication performance metric \cite{kanazawa2025,Bello2025}. The receiver is assumed to select the \gls{CE} and symbol detection method that maximizes $\eta_{\text{eff}}$. For the considered uncoded system, the effective throughput is defined as \cite{Bello2025}
\begin{equation}
\eta_{\text{eff}} = (1 - \text{BER}) \mathcal{D} \log_2 Q,
\end{equation}
where $\text{BER}$ is the bit error rate, $Q$ is the modulation order, and $\mathcal{D}$ is the data density. The data density is given by
\begin{equation}
\mathcal{D} = 
\begin{cases}
1, & \text{SP},\\[1ex]
1 - \mathcal{D}_p = 1 - \frac{1}{K_t K_f}, & \text{EP}.
\end{cases}
\end{equation}
In particular, for conventional \gls{EP} schemes, $\mathcal{D}_p \neq 0$ and thus $\mathcal{D} < 1$ since some \glspl{RE} are reserved for pilots. For \gls{SP} schemes $\mathcal{D} = 1$, as all time-frequency resources carry data \cite{Bello2025}.

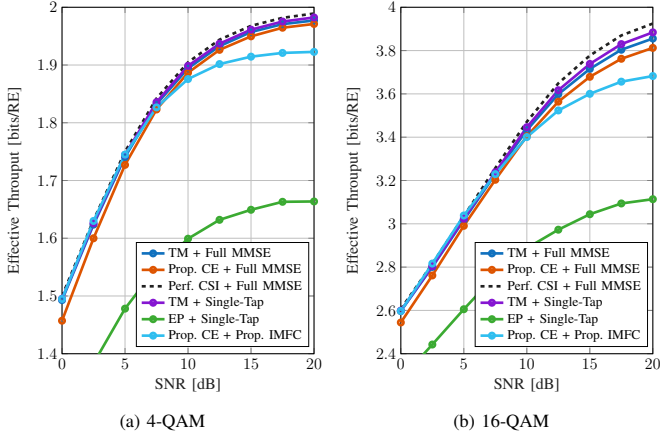
\begin{figure}[t]
 \centering
 \begin{subfigure}[t]{0.24\textwidth} 
 \centering
 \resizebox{\columnwidth}{!}{
%
%
\definecolor{mycolor1}{rgb}{0.06600,0.44300,0.74500}%
\definecolor{mycolor2}{rgb}{0.86600,0.32900,0.00000}%
\definecolor{mycolor3}{rgb}{0.12941,0.12941,0.12941}%
\definecolor{mycolor4}{rgb}{0.52100,0.08600,0.81900}%
\definecolor{mycolor5}{rgb}{0.23100,0.66600,0.19600}%
\definecolor{mycolor6}{rgb}{0.18400,0.74500,0.93700}%
\begin{tikzpicture}

\begin{axis}[%
width=2.61in,
height=3.591in,
at={(0.773in,0.485in)},
scale only axis,
xmin=0,
xmax=20,
xlabel style={font=\color{mycolor3}},
xlabel={\large SNR [dB]},
ymin=1.4,
ymax=2,
ylabel style={font=\color{mycolor3}},
ylabel={\large Effective Throuput [bits/RE]},
axis background/.style={fill=white},
title style={font=\bfseries\color{mycolor3}},
ticklabel style={font=\large},
xmajorgrids,
ymajorgrids,
legend style={at={(0.98,0.02)}, anchor=south east, legend cell align=left, align=left}
]
\addplot [color=mycolor1, line width=2.0pt, mark=o, mark options={solid, mycolor1}]
  table[row sep=crcr]{%
0	1.49286743164062\\
2.5	1.6236572265625\\
5	1.74071044921875\\
7.5	1.83221557617188\\
10	1.894306640625\\
12.5	1.93267333984375\\
15	1.95714111328125\\
17.5	1.97080932617188\\
20	1.97717529296875\\
};
\addlegendentry{TM + Full MMSE}

\addplot [color=mycolor2, line width=2.0pt, mark=o, mark options={solid, mycolor2}]
  table[row sep=crcr]{%
0	1.457177734375\\
2.5	1.59995483398437\\
5	1.72706787109375\\
7.5	1.82304931640625\\
10	1.88672973632813\\
12.5	1.92628784179687\\
15	1.94967529296875\\
17.5	1.9645947265625\\
20	1.97110961914062\\
};
\addlegendentry{Prop. CE + Full MMSE}

\addplot [color=mycolor3, dashed, line width=2.0pt]
  table[row sep=crcr]{%
0	1.49969848632812\\
2.5	1.63134643554687\\
5	1.7499365234375\\
7.5	1.84213989257812\\
10	1.90468017578125\\
12.5	1.94393676757812\\
15	1.96789428710937\\
17.5	1.98171264648438\\
20	1.98918701171875\\
};
\addlegendentry{Perf. CSI + Full MMSE}

\addplot [color=mycolor4, line width=2.0pt, mark=o, mark options={solid, mycolor4}]
  table[row sep=crcr]{%
0	1.49508056640625\\
2.5	1.62606811523438\\
5	1.74431884765625\\
7.5	1.83568969726562\\
10	1.8984765625\\
12.5	1.93695678710937\\
15	1.96126220703125\\
17.5	1.97526733398437\\
20	1.98234619140625\\
};
\addlegendentry{TM + Single-Tap}

\addplot [color=mycolor5, line width=2.0pt, mark=o, mark options={solid, mycolor5}]
  table[row sep=crcr]{%
0	1.2824\\
2.5	1.3833\\
5	1.4781\\
7.5	1.5492\\
10	1.5992\\
12.5	1.6319\\
15	1.6493\\
17.5	1.6631\\
20	1.6637\\
};
\addlegendentry{EP + Single-Tap}

\addplot [color=mycolor6, line width=2.0pt, mark=o, mark options={solid, mycolor6}]
  table[row sep=crcr]{%
0	1.494853515625\\
2.5	1.629951171875\\
5	1.74482666015625\\
7.5	1.82667114257813\\
10	1.87572265625\\
12.5	1.90164794921875\\
15	1.9144775390625\\
17.5	1.92096557617187\\
20	1.92274047851563\\
};
\addlegendentry{Prop. CE + Prop. IMFC}

\end{axis}

\end{tikzpicture}%
}
 \caption{4-QAM}\label{fig:4QAM_TvsSNR_int_speed100}
 \end{subfigure}%
 \hfill 
\begin{subfigure}[t]{0.24\textwidth} 
 \centering
 \resizebox{\columnwidth}{!}{
%
%
\definecolor{mycolor1}{rgb}{0.06600,0.44300,0.74500}%
\definecolor{mycolor2}{rgb}{0.86600,0.32900,0.00000}%
\definecolor{mycolor3}{rgb}{0.12941,0.12941,0.12941}%
\definecolor{mycolor4}{rgb}{0.52100,0.08600,0.81900}%
\definecolor{mycolor5}{rgb}{0.23100,0.66600,0.19600}%
\definecolor{mycolor6}{rgb}{0.18400,0.74500,0.93700}%
\begin{tikzpicture}

\begin{axis}[%
width=2.61in,
height=3.591in,
at={(0.773in,0.485in)},
scale only axis,
xmin=0,
xmax=20,
xlabel style={font=\color{mycolor3}},
xlabel={\large SNR [dB]},
ymin=2.4,
ymax=4,
ylabel style={font=\color{mycolor3}},
ylabel={\large Effective Throuput [bits/RE]},
axis background/.style={fill=white},
title style={font=\bfseries\color{mycolor3}},
ticklabel style={font=\large},
xmajorgrids,
ymajorgrids,
legend style={at={(0.98,0.02)}, anchor=south east, legend cell align=left, align=left}
]
\addplot [color=mycolor1, line width=2.0pt, mark=o, mark options={solid, mycolor1}]
  table[row sep=crcr]{%
0	2.59703979492188\\
2.5	2.80087280273438\\
5	3.01744262695312\\
7.5	3.22985473632813\\
10	3.4323291015625\\
12.5	3.59860473632813\\
15	3.71531127929688\\
17.5	3.80392578125\\
20	3.85558715820313\\
};
\addlegendentry{TM + Full MMSE}

\addplot [color=mycolor2, line width=2.0pt, mark=o, mark options={solid, mycolor2}]
  table[row sep=crcr]{%
0	2.54427368164063\\
2.5	2.76150146484375\\
5	2.990732421875\\
7.5	3.20333374023437\\
10	3.40446899414062\\
12.5	3.56410888671875\\
15	3.67885375976562\\
17.5	3.762138671875\\
20	3.8124267578125\\
};
\addlegendentry{Prop. CE + Full MMSE}

\addplot [color=mycolor3, dashed, line width=2.0pt]
  table[row sep=crcr]{%
0	2.60636474609375\\
2.5	2.81501586914063\\
5	3.0350244140625\\
7.5	3.2575048828125\\
10	3.47119506835937\\
12.5	3.64885009765625\\
15	3.7771044921875\\
17.5	3.8698974609375\\
20	3.92482666015625\\
};
\addlegendentry{Perf. CSI + Full MMSE}

\addplot [color=mycolor4, line width=2.0pt, mark=o, mark options={solid, mycolor4}]
  table[row sep=crcr]{%
0	2.59900634765625\\
2.5	2.80451782226563\\
5	3.02366333007812\\
7.5	3.2378515625\\
10	3.44609375\\
12.5	3.61692260742188\\
15	3.73833129882812\\
17.5	3.829853515625\\
20	3.88406860351563\\
};
\addlegendentry{TM + Single-Tap}

\addplot [color=mycolor5, line width=2.0pt, mark=o, mark options={solid, mycolor5}]
  table[row sep=crcr]{%
0	2.2817\\
2.5	2.4425\\
5	2.6055\\
7.5	2.7614\\
10	2.8874\\
12.5	2.9728\\
15	3.0442\\
17.5	3.094\\
20	3.1137\\
};
\addlegendentry{EP + Single-Tap}

\addplot [color=mycolor6, line width=2.0pt, mark=o, mark options={solid, mycolor6}]
  table[row sep=crcr]{%
0	2.59740234375\\
2.5	2.81646728515625\\
5	3.03841430664062\\
7.5	3.229423828125\\
10	3.40055419921875\\
12.5	3.52356811523437\\
15	3.60028686523437\\
17.5	3.65644409179688\\
20	3.68231079101562\\
};
\addlegendentry{Prop. CE + Prop. IMFC}

\end{axis}

\end{tikzpicture}%
}
 \caption{16-QAM}\label{fig:16QAM_TvsSNR_int_speed100}
 \end{subfigure}%
 \caption{Effective throughput against SNR for a fixed PDR of $30$ dB. The channel has integer delays and Doppler shifts and the maximum speed is $100$ km/h.} \label{fig:int_lowspeed}
\end{figure}

\begin{figure}[t]
 \centering
 \begin{subfigure}[t]{0.24\textwidth} 
 \centering
 \resizebox{\columnwidth}{!}{
%
%
\definecolor{mycolor1}{rgb}{0.06600,0.44300,0.74500}%
\definecolor{mycolor2}{rgb}{0.86600,0.32900,0.00000}%
\definecolor{mycolor3}{rgb}{0.12941,0.12941,0.12941}%
\definecolor{mycolor4}{rgb}{0.52100,0.08600,0.81900}%
\definecolor{mycolor5}{rgb}{0.23100,0.66600,0.19600}%
\definecolor{mycolor6}{rgb}{0.18400,0.74500,0.93700}%
\begin{tikzpicture}

\begin{axis}[%
width=2.61in,
height=3.591in,
at={(0.773in,0.485in)},
scale only axis,
xmin=0,
xmax=20,
xlabel style={font=\color{mycolor3}},
xlabel={\large SNR [dB]},
ymin=1.4,
ymax=2,
ylabel style={font=\color{mycolor3}},
ylabel={\large Effective Throuput [bits/RE]},
axis background/.style={fill=white},
title style={font=\bfseries\color{mycolor3}},
ticklabel style={font=\large},
xmajorgrids,
ymajorgrids,
legend style={at={(0.98,0.02)}, anchor=south east, legend cell align=left, align=left}
]
\addplot [color=mycolor1, line width=2.0pt, mark=o, mark options={solid, mycolor1}]
  table[row sep=crcr]{%
0	1.4054736328125\\
2.5	1.52129516601563\\
5	1.62773071289062\\
7.5	1.7033056640625\\
10	1.7540771484375\\
12.5	1.78398559570312\\
15	1.8084765625\\
17.5	1.81890380859375\\
20	1.82377807617187\\
};
\addlegendentry{TM + Full MMSE}

\addplot [color=mycolor2, line width=2.0pt, mark=o, mark options={solid, mycolor2}]
  table[row sep=crcr]{%
0	1.45411865234375\\
2.5	1.593046875\\
5	1.72498901367188\\
7.5	1.8218408203125\\
10	1.88275390625\\
12.5	1.92156372070313\\
15	1.94558471679687\\
17.5	1.960810546875\\
20	1.96929931640625\\
};
\addlegendentry{Prop. CE + Full MMSE}

\addplot [color=mycolor3, dashed, line width=2.0pt]
  table[row sep=crcr]{%
0	1.50027587890625\\
2.5	1.62905395507812\\
5	1.7502001953125\\
7.5	1.84438232421875\\
10	1.9047412109375\\
12.5	1.94405639648437\\
15	1.96746337890625\\
17.5	1.98163452148437\\
20	1.990048828125\\
};
\addlegendentry{Perf. CSI + Full MMSE}

\addplot [color=mycolor4, line width=2.0pt, mark=o, mark options={solid, mycolor4}]
  table[row sep=crcr]{%
0	1.40709594726563\\
2.5	1.52365478515625\\
5	1.63039794921875\\
7.5	1.70655883789063\\
10	1.75811889648438\\
12.5	1.7882373046875\\
15	1.81243530273437\\
17.5	1.82372680664062\\
20	1.82853515625\\
};
\addlegendentry{TM + Single-Tap}

\addplot [color=mycolor5, line width=2.0pt, mark=o, mark options={solid, mycolor5}]
  table[row sep=crcr]{%
0	1.2811\\
2.5	1.3814\\
5	1.4692\\
7.5	1.5422\\
10	1.5946\\
12.5	1.62575\\
15	1.6437\\
17.5	1.6554\\
20	1.6611\\
};
\addlegendentry{EP + Single-Tap}

\addplot [color=mycolor6, line width=2.0pt, mark=o, mark options={solid, mycolor6}]
  table[row sep=crcr]{%
0	1.49004272460938\\
2.5	1.62313842773438\\
5	1.74286499023437\\
7.5	1.82556640625\\
10	1.8723291015625\\
12.5	1.89797973632812\\
15	1.91066040039062\\
17.5	1.9194482421875\\
20	1.92296997070312\\
};
\addlegendentry{Prop. CE + Prop. IMFC}

\end{axis}

\end{tikzpicture}%
}
 \caption{4-QAM}\label{fig:4QAM_TvsSNR_frac_speed100}
 \end{subfigure}%
 \hfill 
\begin{subfigure}[t]{0.24\textwidth} 
 \centering
 \resizebox{\columnwidth}{!}{
%
%
\definecolor{mycolor1}{rgb}{0.06600,0.44300,0.74500}%
\definecolor{mycolor2}{rgb}{0.86600,0.32900,0.00000}%
\definecolor{mycolor3}{rgb}{0.12941,0.12941,0.12941}%
\definecolor{mycolor4}{rgb}{0.52100,0.08600,0.81900}%
\definecolor{mycolor5}{rgb}{0.23100,0.66600,0.19600}%
\definecolor{mycolor6}{rgb}{0.18400,0.74500,0.93700}%
\begin{tikzpicture}

\begin{axis}[%
width=2.61in,
height=3.591in,
at={(0.773in,0.485in)},
scale only axis,
xmin=0,
xmax=20,
xlabel style={font=\color{mycolor3}},
xlabel={\large SNR [dB]},
ymin=2.4,
ymax=4,
ylabel style={font=\color{mycolor3}},
ylabel={\large Effective Throuput [bits/RE]},
axis background/.style={fill=white},
title style={font=\bfseries\color{mycolor3}},
ticklabel style={font=\large},
xmajorgrids,
ymajorgrids,
legend style={at={(0.98,0.02)}, anchor=south east, legend cell align=left, align=left}
]
\addplot [color=mycolor1, line width=2.0pt, mark=o, mark options={solid, mycolor1}]
  table[row sep=crcr]{%
0	2.48334594726562\\
2.5	2.66660766601562\\
5	2.83517456054688\\
7.5	2.99155029296875\\
10	3.12170532226563\\
12.5	3.20884521484375\\
15	3.26564331054688\\
17.5	3.31566772460938\\
20	3.34840454101563\\
};
\addlegendentry{TM + Full MMSE}

\addplot [color=mycolor2, line width=2.0pt, mark=o, mark options={solid, mycolor2}]
  table[row sep=crcr]{%
0	2.5405322265625\\
2.5	2.75851318359375\\
5	2.98771606445313\\
7.5	3.20094848632812\\
10	3.39651733398438\\
12.5	3.553203125\\
15	3.66778930664062\\
17.5	3.74268676757813\\
20	3.794638671875\\
};
\addlegendentry{Prop. CE + Full MMSE}

\addplot [color=mycolor3, dashed, line width=2.0pt]
  table[row sep=crcr]{%
0	2.60685668945313\\
2.5	2.81582885742188\\
5	3.03533203125\\
7.5	3.26105102539062\\
10	3.473828125\\
12.5	3.64969360351563\\
15	3.7827783203125\\
17.5	3.86884887695313\\
20	3.92593505859375\\
};
\addlegendentry{Perf. CSI + Full MMSE}

\addplot [color=mycolor4, line width=2.0pt, mark=o, mark options={solid, mycolor4}]
  table[row sep=crcr]{%
0	2.48569091796875\\
2.5	2.66923828125\\
5	2.84004272460937\\
7.5	2.99709350585938\\
10	3.12940307617187\\
12.5	3.21705078125\\
15	3.27529296875\\
17.5	3.32673583984375\\
20	3.36166625976563\\
};
\addlegendentry{TM + Single-Tap}

\addplot [color=mycolor5, line width=2.0pt, mark=o, mark options={solid, mycolor5}]
  table[row sep=crcr]{%
0	2.2791\\
2.5	2.4406\\
5	2.6015\\
7.5	2.7546\\
10	2.8741\\
12.5	2.9675\\
15	3.0274\\
17.5	3.0683\\
20	3.0965\\
};
\addlegendentry{EP + Single-Tap}

\addplot [color=mycolor6, line width=2.0pt, mark=o, mark options={solid, mycolor6}]
  table[row sep=crcr]{%
0	2.5930517578125\\
2.5	2.81181396484375\\
5	3.0331298828125\\
7.5	3.22536376953125\\
10	3.39232666015625\\
12.5	3.5126171875\\
15	3.595673828125\\
17.5	3.6390771484375\\
20	3.67266235351563\\
};
\addlegendentry{Prop. CE + Prop. IMFC}

\end{axis}

\end{tikzpicture}%
}
 \caption{16-QAM}\label{fig:16QAM_TvsSNR_frac_speed100}
 \end{subfigure}%
 \caption{Effective throughput against SNR for a fixed PDR of $30$ dB. The channel has fractional delays and Doppler shifts and the maximum speed is $100$ km/h.} \label{fig:frac_lowspeed}
\end{figure}
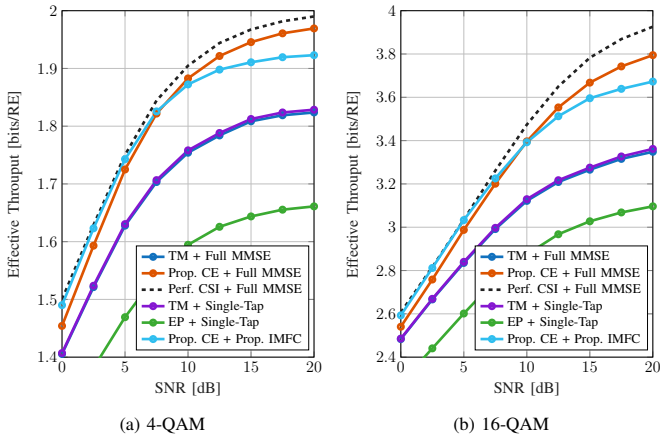

\subsection{Optimum PDR Selection}
Given the \gls{SP} scheme discussed in Section \ref{sec:SPscheme}, it can be noted that the \gls{PDR} governs the trade-off between \gls{CE} accuracy and data detection quality. For a fixed transmit power $P_T$, increasing the \gls{PDR} boosts the pilot amplitude, improving channel estimation \cite{Bello2025}, but reduces the power allocated to data symbols, thereby degrading the detection \gls{SNR}. Conversely, lowering the \gls{PDR} enhances the detection \gls{SNR} but reduces pilot power, leading to poorer \gls{CE} accuracy.  
This intrinsic trade-off results in an optimal \gls{PDR} that maximizes effective throughput. Figure \ref{fig:TvsPDR} illustrates that all delay-Doppler \gls{SP}-based methods achieve their highest $\eta_{\text{eff}}$ at specific \gls{PDR} values. In particular, the optimum \gls{PDR} that maximizes effective throughput is approximately $30$ dB. Deviating from this value, either by increasing or decreasing the \gls{PDR}, causes $\eta_{\text{eff}}$ to drop, eventually falling below that of the conventional EP + Single-Tap scheme.

\begin{figure}[t]
 \centering
 \begin{subfigure}[t]{0.24\textwidth} 
 \centering
 \resizebox{\columnwidth}{!}{
%
%
\definecolor{mycolor1}{rgb}{0.06600,0.44300,0.74500}%
\definecolor{mycolor2}{rgb}{0.86600,0.32900,0.00000}%
\definecolor{mycolor3}{rgb}{0.12941,0.12941,0.12941}%
\definecolor{mycolor4}{rgb}{0.52100,0.08600,0.81900}%
\definecolor{mycolor5}{rgb}{0.23100,0.66600,0.19600}%
\definecolor{mycolor6}{rgb}{0.18400,0.74500,0.93700}%
\begin{tikzpicture}

\begin{axis}[%
width=2.61in,
height=3.591in,
at={(0.773in,0.485in)},
scale only axis,
xmin=0,
xmax=20,
xlabel style={font=\color{mycolor3}},
xlabel={\large SNR [dB]},
ymin=1,
ymax=2,
ylabel style={font=\color{mycolor3}},
ylabel={\large Effective Throuput [bits/RE]},
axis background/.style={fill=white},
title style={font=\bfseries\color{mycolor3}},
ticklabel style={font=\large},
xmajorgrids,
ymajorgrids,
legend style={at={(0.98,0.02)}, anchor=south east, legend cell align=left, align=left}
]
\addplot [color=mycolor1, line width=2.0pt, mark=o, mark options={solid, mycolor1}]
  table[row sep=crcr]{%
0	1.39361206054688\\
2.5	1.49501831054687\\
5	1.57702026367188\\
7.5	1.636376953125\\
10	1.66998046875\\
12.5	1.692109375\\
15	1.70043334960938\\
17.5	1.714765625\\
20	1.724091796875\\
};
\addlegendentry{TM + Full MMSE}

\addplot [color=mycolor2, line width=2.0pt, mark=o, mark options={solid, mycolor2}]
  table[row sep=crcr]{%
0	1.45936767578125\\
2.5	1.59942138671875\\
5	1.72871215820313\\
7.5	1.82416381835937\\
10	1.886806640625\\
12.5	1.92641723632812\\
15	1.95068603515625\\
17.5	1.9633740234375\\
20	1.9720947265625\\
};
\addlegendentry{Prop. CE + Full MMSE}

\addplot [color=mycolor3, dashed, line width=2.0pt]
  table[row sep=crcr]{%
0	1.50322875976563\\
2.5	1.6320068359375\\
5	1.75133544921875\\
7.5	1.84390380859375\\
10	1.9052197265625\\
12.5	1.94403564453125\\
15	1.96857299804687\\
17.5	1.98154907226563\\
20	1.9899609375\\
};
\addlegendentry{Perf. CSI + Full MMSE}

\addplot [color=mycolor4, line width=2.0pt, mark=o, mark options={solid, mycolor4}]
  table[row sep=crcr]{%
0	1.472255859375\\
2.5	1.5948681640625\\
5	1.70529663085938\\
7.5	1.78982666015625\\
10	1.84466674804688\\
12.5	1.88071044921875\\
15	1.90121337890625\\
17.5	1.91614990234375\\
20	1.92598388671875\\
};
\addlegendentry{TM + Single-Tap}

\addplot [color=mycolor5, line width=2.0pt, mark=o, mark options={solid, mycolor5}]
  table[row sep=crcr]{%
0	1.0666\\
2.5	1.1045\\
5	1.1333\\
7.5	1.1613\\
10	1.1771\\
12.5	1.1913\\
15	1.1942\\
17.5	1.2039\\
20	1.2166\\
};
\addlegendentry{EP + Single-Tap}

\addplot [color=mycolor6, line width=2.0pt, mark=o, mark options={solid, mycolor6}]
  table[row sep=crcr]{%
0	1.49303100585937\\
2.5	1.62739379882813\\
5	1.7463916015625\\
7.5	1.8293115234375\\
10	1.87755615234375\\
12.5	1.90407104492187\\
15	1.91796875\\
17.5	1.92345092773437\\
20	1.92932495117188\\
};
\addlegendentry{Prop. CE + Prop. IMFC}

\end{axis}

\end{tikzpicture}%
}
 \caption{4-QAM}\label{fig:4QAM_TvsSNR_int_speed1e3}
 \end{subfigure}%
 \hfill 
\begin{subfigure}[t]{0.24\textwidth} 
 \centering
 \resizebox{\columnwidth}{!}{
%
%
\definecolor{mycolor1}{rgb}{0.06600,0.44300,0.74500}%
\definecolor{mycolor2}{rgb}{0.86600,0.32900,0.00000}%
\definecolor{mycolor3}{rgb}{0.12941,0.12941,0.12941}%
\definecolor{mycolor4}{rgb}{0.52100,0.08600,0.81900}%
\definecolor{mycolor5}{rgb}{0.23100,0.66600,0.19600}%
\definecolor{mycolor6}{rgb}{0.18400,0.74500,0.93700}%
\begin{tikzpicture}

\begin{axis}[%
width=2.61in,
height=3.591in,
at={(0.773in,0.485in)},
scale only axis,
xmin=0,
xmax=20,
xlabel style={font=\color{mycolor3}},
xlabel={\large SNR [dB]},
ymin=2,
ymax=4,
ylabel style={font=\color{mycolor3}},
ylabel={\large Effective Throuput [bits/RE]},
axis background/.style={fill=white},
title style={font=\bfseries\color{mycolor3}},
ticklabel style={font=\large},
xmajorgrids,
ymajorgrids,
legend style={at={(0.98,0.02)}, anchor=south east, legend cell align=left, align=left}
]
\addplot [color=mycolor1, line width=2.0pt, mark=o, mark options={solid, mycolor1}]
  table[row sep=crcr]{%
0	2.46923217773438\\
2.5	2.611064453125\\
5	2.74119873046875\\
7.5	2.83326171875\\
10	2.94757690429687\\
12.5	2.9993310546875\\
15	3.02215576171875\\
17.5	3.01189331054688\\
20	3.02428100585937\\
};
\addlegendentry{TM + Full MMSE}

\addplot [color=mycolor2, line width=2.0pt, mark=o, mark options={solid, mycolor2}]
  table[row sep=crcr]{%
0	2.54204711914063\\
2.5	2.75539306640625\\
5	2.98414672851562\\
7.5	3.19861572265625\\
10	3.3914990234375\\
12.5	3.55282592773438\\
15	3.6649853515625\\
17.5	3.74265258789063\\
20	3.79295654296875\\
};
\addlegendentry{Prop. CE + Full MMSE}

\addplot [color=mycolor3, dashed, line width=2.0pt]
  table[row sep=crcr]{%
0	2.60578002929687\\
2.5	2.8082568359375\\
5	3.02840942382812\\
7.5	3.25171997070313\\
10	3.45692504882813\\
12.5	3.63626342773438\\
15	3.76426879882812\\
17.5	3.85753051757812\\
20	3.91149780273438\\
};
\addlegendentry{Perf. CSI + Full MMSE}

\addplot [color=mycolor4, line width=2.0pt, mark=o, mark options={solid, mycolor4}]
  table[row sep=crcr]{%
0	2.56727783203125\\
2.5	2.75158081054688\\
5	2.94282592773437\\
7.5	3.11807861328125\\
10	3.28461181640625\\
12.5	3.40658569335938\\
15	3.48492797851562\\
17.5	3.52739501953125\\
20	3.5640283203125\\
};
\addlegendentry{TM + Single-Tap}

\addplot [color=mycolor5, line width=2.0pt, mark=o, mark options={solid, mycolor5}]
  table[row sep=crcr]{%
0	2.0617\\
2.5	2.1369\\
5	2.2143\\
7.5	2.272\\
10	2.3187\\
12.5	2.3451\\
15	2.3755\\
17.5	2.3572\\
20	2.3628\\
};
\addlegendentry{EP + Single-Tap}

\addplot [color=mycolor6, line width=2.0pt, mark=o, mark options={solid, mycolor6}]
  table[row sep=crcr]{%
0	2.5894287109375\\
2.5	2.8074951171875\\
5	3.0319580078125\\
7.5	3.22711669921875\\
10	3.3877197265625\\
12.5	3.51308715820312\\
15	3.58908081054688\\
17.5	3.6396826171875\\
20	3.66322875976562\\
};
\addlegendentry{Prop. CE + Prop. IMFC}

\end{axis}

\end{tikzpicture}%
}
 \caption{16-QAM}\label{fig:16QAM_TvsSNR_int_speed1e3}
 \end{subfigure}%

 \caption{Effective throughput against SNR for a fixed PDR of $30$ dB. The channel has integer delays and Doppler shifts and the maximum speed is $1000$ km/h.} \label{fig:int_highspeed}
\end{figure}
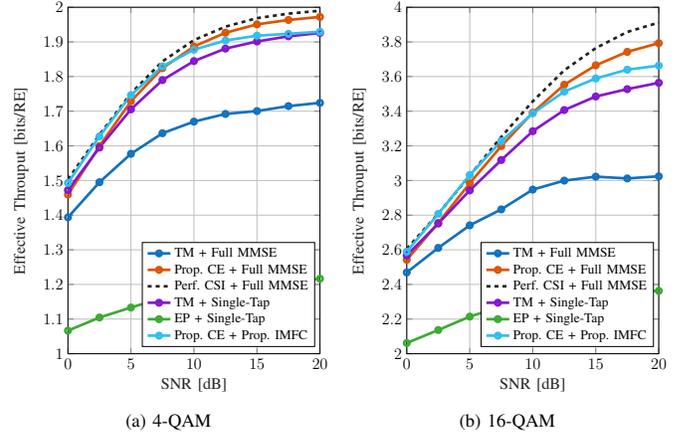

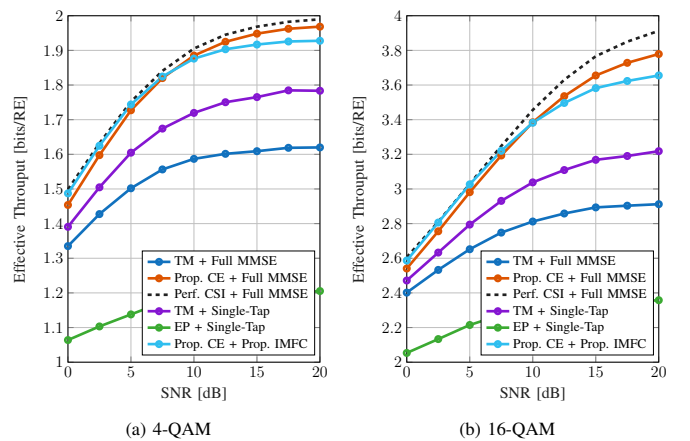
\begin{figure}[t]
 \centering
 \begin{subfigure}[t]{0.24\textwidth} 
 \centering
 \resizebox{\columnwidth}{!}{
%
%
\definecolor{mycolor1}{rgb}{0.06600,0.44300,0.74500}%
\definecolor{mycolor2}{rgb}{0.86600,0.32900,0.00000}%
\definecolor{mycolor3}{rgb}{0.12941,0.12941,0.12941}%
\definecolor{mycolor4}{rgb}{0.52100,0.08600,0.81900}%
\definecolor{mycolor5}{rgb}{0.23100,0.66600,0.19600}%
\definecolor{mycolor6}{rgb}{0.18400,0.74500,0.93700}%
\begin{tikzpicture}

\begin{axis}[%
width=2.61in,
height=3.591in,
at={(0.773in,0.485in)},
scale only axis,
xmin=0,
xmax=20,
xlabel style={font=\color{mycolor3}},
xlabel={\large SNR [dB]},
ymin=1,
ymax=2,
ylabel style={font=\color{mycolor3}},
ylabel={\large Effective Throuput [bits/RE]},
axis background/.style={fill=white},
title style={font=\bfseries\color{mycolor3}},
ticklabel style={font=\large},
xmajorgrids,
ymajorgrids,
legend style={at={(0.98,0.02)}, anchor=south east, legend cell align=left, align=left}
]
\addplot [color=mycolor1, line width=2.0pt, mark=o, mark options={solid, mycolor1}]
  table[row sep=crcr]{%
0	1.33503662109375\\
2.5	1.42749145507813\\
5	1.50197998046875\\
7.5	1.55638061523437\\
10	1.58692016601562\\
12.5	1.60143188476563\\
15	1.60903930664062\\
17.5	1.61880737304688\\
20	1.61987060546875\\
};
\addlegendentry{TM + Full MMSE}

\addplot [color=mycolor2, line width=2.0pt, mark=o, mark options={solid, mycolor2}]
  table[row sep=crcr]{%
0	1.45381591796875\\
2.5	1.59758178710937\\
5	1.72717407226563\\
7.5	1.82003173828125\\
10	1.88502685546875\\
12.5	1.92480102539062\\
15	1.94831298828125\\
17.5	1.9623974609375\\
20	1.96848022460938\\
};
\addlegendentry{Prop. CE + Full MMSE}

\addplot [color=mycolor3, dashed, line width=2.0pt]
  table[row sep=crcr]{%
0	1.4988232421875\\
2.5	1.63206298828125\\
5	1.751845703125\\
7.5	1.84086303710938\\
10	1.9053955078125\\
12.5	1.94517822265625\\
15	1.96843627929688\\
17.5	1.9824853515625\\
20	1.98966674804688\\
};
\addlegendentry{Perf. CSI + Full MMSE}

\addplot [color=mycolor4, line width=2.0pt, mark=o, mark options={solid, mycolor4}]
  table[row sep=crcr]{%
0	1.390732421875\\
2.5	1.50472900390625\\
5	1.60476440429687\\
7.5	1.67425537109375\\
10	1.71957763671875\\
12.5	1.75037353515625\\
15	1.76513427734375\\
17.5	1.78463500976562\\
20	1.78343017578125\\
};
\addlegendentry{TM + Single-Tap}

\addplot [color=mycolor5, line width=2.0pt, mark=o, mark options={solid, mycolor5}]
  table[row sep=crcr]{%
0	1.0639\\
2.5	1.1032\\
5	1.1379\\
7.5	1.1728\\
10	1.1765\\
12.5	1.1911\\
15	1.1953\\
17.5	1.1983\\
20	1.2053\\
};
\addlegendentry{EP + Single-Tap}

\addplot [color=mycolor6, line width=2.0pt, mark=o, mark options={solid, mycolor6}]
  table[row sep=crcr]{%
0	1.487265625\\
2.5	1.62477172851563\\
5	1.74381713867188\\
7.5	1.82447021484375\\
10	1.87620727539062\\
12.5	1.90327270507813\\
15	1.916884765625\\
17.5	1.92571655273437\\
20	1.92769653320313\\
};
\addlegendentry{Prop. CE + Prop. IMFC}

\end{axis}

\end{tikzpicture}%
}
 \caption{4-QAM}\label{fig:4QAM_TvsSNR_frac_speed1e3}
 \end{subfigure}%
 \hfill 
\begin{subfigure}[t]{0.24\textwidth} 
 \centering
 \resizebox{\columnwidth}{!}{
%
%
\definecolor{mycolor1}{rgb}{0.06600,0.44300,0.74500}%
\definecolor{mycolor2}{rgb}{0.86600,0.32900,0.00000}%
\definecolor{mycolor3}{rgb}{0.12941,0.12941,0.12941}%
\definecolor{mycolor4}{rgb}{0.52100,0.08600,0.81900}%
\definecolor{mycolor5}{rgb}{0.23100,0.66600,0.19600}%
\definecolor{mycolor6}{rgb}{0.18400,0.74500,0.93700}%
\begin{tikzpicture}

\begin{axis}[%
width=2.61in,
height=3.591in,
at={(0.773in,0.485in)},
scale only axis,
xmin=0,
xmax=20,
xlabel style={font=\color{mycolor3}},
xlabel={\large SNR [dB]},
ymin=2,
ymax=4,
ylabel style={font=\color{mycolor3}},
ylabel={\large Effective Throuput [bits/RE]},
axis background/.style={fill=white},
title style={font=\bfseries\color{mycolor3}},
ticklabel style={font=\large},
xmajorgrids,
ymajorgrids,
legend style={at={(0.98,0.02)}, anchor=south east, legend cell align=left, align=left}
]
\addplot [color=mycolor1, line width=2.0pt, mark=o, mark options={solid, mycolor1}]
  table[row sep=crcr]{%
0	2.401708984375\\
2.5	2.53277099609375\\
5	2.65210571289062\\
7.5	2.74810424804688\\
10	2.81208374023438\\
12.5	2.8583203125\\
15	2.89364990234375\\
17.5	2.90342529296875\\
20	2.91212158203125\\
};
\addlegendentry{TM + Full MMSE}

\addplot [color=mycolor2, line width=2.0pt, mark=o, mark options={solid, mycolor2}]
  table[row sep=crcr]{%
0	2.54083251953125\\
2.5	2.75578979492188\\
5	2.98141967773438\\
7.5	3.19323486328125\\
10	3.38482299804688\\
12.5	3.53568115234375\\
15	3.655146484375\\
17.5	3.72821533203125\\
20	3.778994140625\\
};
\addlegendentry{Prop. CE + Full MMSE}

\addplot [color=mycolor3, dashed, line width=2.0pt]
  table[row sep=crcr]{%
0	2.60625732421875\\
2.5	2.81070068359375\\
5	3.0293408203125\\
7.5	3.24799926757813\\
10	3.456025390625\\
12.5	3.6303515625\\
15	3.76629760742187\\
17.5	3.85032836914062\\
20	3.91280151367187\\
};
\addlegendentry{Perf. CSI + Full MMSE}

\addplot [color=mycolor4, line width=2.0pt, mark=o, mark options={solid, mycolor4}]
  table[row sep=crcr]{%
0	2.47156860351563\\
2.5	2.63278564453125\\
5	2.79439697265625\\
7.5	2.93123168945313\\
10	3.0380029296875\\
12.5	3.109208984375\\
15	3.16826904296875\\
17.5	3.19032348632813\\
20	3.21810302734375\\
};
\addlegendentry{TM + Single-Tap}

\addplot [color=mycolor5, line width=2.0pt, mark=o, mark options={solid, mycolor5}]
  table[row sep=crcr]{%
0	2.0534\\
2.5	2.1334\\
5	2.2148\\
7.5	2.2738\\
10	2.32\\
12.5	2.3326\\
15	2.3535\\
17.5	2.3463\\
20	2.3578\\
};
\addlegendentry{EP + Single-Tap}

\addplot [color=mycolor6, line width=2.0pt, mark=o, mark options={solid, mycolor6}]
  table[row sep=crcr]{%
0	2.58693359375\\
2.5	2.80616333007813\\
5	3.02667114257812\\
7.5	3.22015991210937\\
10	3.38217651367188\\
12.5	3.49650390625\\
15	3.58214477539063\\
17.5	3.62338989257813\\
20	3.65517944335937\\
};
\addlegendentry{Prop. CE + Prop. IMFC}

\end{axis}

\end{tikzpicture}%
}
 \caption{16-QAM}\label{fig:16QAM_TvsSNR_frac_speed1e3}
 \end{subfigure}%

 \caption{Effective throughput against SNR for a fixed PDR of $30$ dB. The channel has fractional delays and Doppler shifts and the maximum speed is $1000$ km/h.} \label{fig:frac_highspeed}
\end{figure}

\subsection{Effective Throughput vs SNR}
The communication performance is evaluated for both integer and fractional channel models. Figures \ref{fig:int_lowspeed} and \ref{fig:frac_lowspeed} show the effective throughput for $4$-QAM and $16$-QAM under integer and fractional channel parameters, respectively. The maximum speed is $100$ km/h, so that \gls{ICI} is negligible. 
It can be observed that delay-Doppler \gls{SP} schemes consistently outperform the conventional \gls{EP}-based scheme. In particular, Figure \ref{fig:int_lowspeed} shows that the TM + Single-Tap method \cite{Bello2025} achieves performance close to the Perf. CSI + Full MMSE bound when channel parameters are integers. In contrast, for the more realistic fractional channel in Figure \ref{fig:frac_lowspeed}, the TM + Single-Tap performance degrades significantly due to the violation of the integer assumption. The proposed Prop. CE + Prop. IMFC scheme, instead, achieves performance close to the Perf. CSI + Full MMSE bound in both integer and fractional scenarios, outperforming the TM + Single-Tap baseline when fractional channel parameters are present. Furthermore, in both Figures \ref{fig:int_lowspeed} and \ref{fig:frac_lowspeed}, the Prop. CE + Prop. IMFC has comparable performance to that of the Prop. CE + Full MMSE, except for a small drop at high \gls{SNR} values, indicating that the proposed iterative detection scheme achieves high effective throughput at much lower computational complexity (linear in $M$ versus cubic for full MMSE).

Figures \ref{fig:int_highspeed} and \ref{fig:frac_highspeed} depict the performance at a higher mobility scenario with maximum speed of $1000$ km/h, representative of extreme 6G conditions. In this case, the Prop. CE + Prop. IMFC outperforms the TM + Single-Tap in both integer and fractional channels, as \gls{ICI} becomes non-negligible. The TM + Single-Tap method fails to compensate for \gls{ICI}, while the proposed approach accounts for both \gls{ICI} and fractional delays/Doppler shifts, maintaining robust performance. The performance gap is especially pronounced for fractional channels, highlighting the advantage of the proposed method under realistic conditions. Overall, Prop. CE + Prop. IMFC remains closer to the Perf. CSI + Full MMSE bound across all considered scenarios.

Interestingly, Figures \ref{fig:int_highspeed} and \ref{fig:frac_highspeed} also show that TM + Single-Tap outperforms TM + Full MMSE despite \gls{ICI} being non-negligible. This is because the TM-based \gls{CE} method does not model \gls{ICI}, while the full MMSE equalizer attempts to compensate for it. The mismatch leads to suboptimal performance for TM + Full MMSE, whereas the \gls{ICI}-free TM + Single-Tap yields better throughput under these conditions.

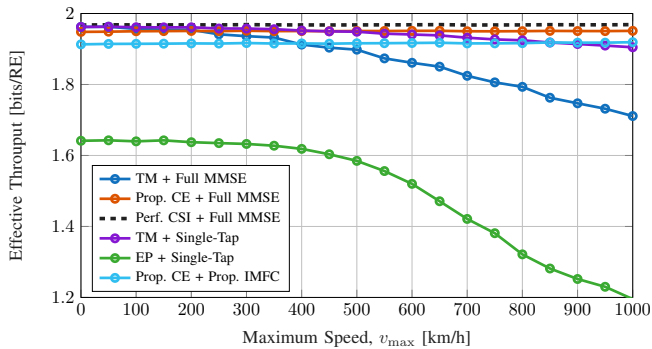
\begin{figure}[t]
 \centering
 \resizebox{0.99\columnwidth}{!}{
%
%
\definecolor{mycolor1}{rgb}{0.06600,0.44300,0.74500}%
\definecolor{mycolor2}{rgb}{0.86600,0.32900,0.00000}%
\definecolor{mycolor3}{rgb}{0.12941,0.12941,0.12941}%
\definecolor{mycolor4}{rgb}{0.52100,0.08600,0.81900}%
\definecolor{mycolor5}{rgb}{0.23100,0.66600,0.19600}%
\definecolor{mycolor6}{rgb}{0.18400,0.74500,0.93700}%
\begin{tikzpicture}

\begin{axis}[%
width=4.225in,
height=2.176in,
scale only axis,
xmin=0,
xmax=1000,
xticklabel style={
    /pgf/number format/set thousands separator={},
    /pgf/number format/1000 sep={}, 
},
xlabel style={font=\color{mycolor3}},
xlabel={Maximum Speed, $v_{\max}$ [km/h]},
ymin=1.2,
ymax=2,
ylabel style={font=\color{mycolor3}},
ylabel={Effective Throuput [bits/RE]},
axis background/.style={fill=white},
title style={font=\bfseries\color{mycolor3}},
xmajorgrids,
ymajorgrids,
legend style={at={(0.02,0.02)}, anchor=south west, legend cell align=left, align=left,font=\footnotesize}
]
\addplot [color=mycolor1, line width=1.5pt, mark=o, mark options={solid, mycolor1}]
  table[row sep=crcr]{%
0	1.96215380859375\\
50	1.9627265625\\
100	1.95642944335937\\
150	1.95452783203125\\
200	1.95401147460938\\
250	1.94130590820312\\
300	1.93613623046875\\
350	1.93220922851562\\
400	1.9124375\\
450	1.90373779296875\\
500	1.89811157226562\\
550	1.87354541015625\\
600	1.86116796875\\
650	1.85023803710937\\
700	1.82440795898437\\
750	1.80598291015625\\
800	1.793365234375\\
850	1.76267138671875\\
900	1.74699243164063\\
950	1.731705078125\\
1000	1.71109375\\
};
\addlegendentry{TM + Full MMSE}

\addplot [color=mycolor2, line width=1.5pt, mark=o, mark options={solid, mycolor2}]
  table[row sep=crcr]{%
0	1.94819921875\\
50	1.94879956054688\\
100	1.9499912109375\\
150	1.95040014648437\\
200	1.950869140625\\
250	1.95038012695312\\
300	1.951427734375\\
350	1.95031591796875\\
400	1.9505859375\\
450	1.95011889648437\\
500	1.9506787109375\\
550	1.950642578125\\
600	1.95110986328125\\
650	1.95127172851562\\
700	1.94996801757813\\
750	1.94977124023437\\
800	1.95023120117187\\
850	1.95097973632812\\
900	1.95057861328125\\
950	1.95041650390625\\
1000	1.95108715820313\\
};
\addlegendentry{Prop. CE + Full MMSE}

\addplot [color=mycolor3, dashed, line width=2pt]
  table[row sep=crcr]{%
0	1.96745288085937\\
50	1.96822802734375\\
100	1.96785595703125\\
150	1.9678359375\\
200	1.96809887695312\\
250	1.9678447265625\\
300	1.96862280273438\\
350	1.96797778320312\\
400	1.96782421875\\
450	1.96768286132812\\
500	1.96786108398438\\
550	1.96801953125\\
600	1.96821459960937\\
650	1.96866967773438\\
700	1.96759521484375\\
750	1.96770483398438\\
800	1.96789086914062\\
850	1.96852001953125\\
900	1.96823559570313\\
950	1.9680947265625\\
1000	1.96866821289062\\
};
\addlegendentry{Perf. CSI + Full MMSE}

\addplot [color=mycolor4, line width=1.5pt, mark=o, mark options={solid, mycolor4}]
  table[row sep=crcr]{%
0	1.96212890625\\
50	1.96275146484375\\
100	1.96121557617188\\
150	1.960927734375\\
200	1.9610888671875\\
250	1.95778100585937\\
300	1.95761303710937\\
350	1.95583251953125\\
400	1.95195922851563\\
450	1.94969897460937\\
500	1.94866430664063\\
550	1.94315307617187\\
600	1.9408310546875\\
650	1.93836572265625\\
700	1.9317587890625\\
750	1.92700219726563\\
800	1.92450903320312\\
850	1.91816625976563\\
900	1.91405322265625\\
950	1.90959204101563\\
1000	1.90483618164062\\
};
\addlegendentry{TM + Single-Tap}

\addplot [color=mycolor5, line width=1.5pt, mark=o, mark options={solid, mycolor5}]
  table[row sep=crcr]{%
0	1.6414\\
50	1.6427\\
100	1.6398\\
150	1.6426\\
200	1.6374\\
250	1.6348\\
300	1.6325\\
350	1.6275\\
400	1.6186\\
450	1.6034\\
500	1.5848\\
550	1.5561\\
600	1.5200\\
650	1.4710\\
700	1.4212\\
750	1.3810\\
800	1.3214\\
850	1.2815\\
900	1.2520\\
950	1.2300\\
1000	1.1953\\
};
\addlegendentry{EP + Single-Tap}

\addplot [color=mycolor6, line width=1.5pt, mark=o, mark options={solid, mycolor6}]
  table[row sep=crcr]{%
0	1.91303662109375\\
50	1.9143095703125\\
100	1.91458544921875\\
150	1.91500219726563\\
200	1.91575122070313\\
250	1.91530078125\\
300	1.91678369140625\\
350	1.91538696289063\\
400	1.91569287109375\\
450	1.91494921875\\
500	1.91604931640625\\
550	1.916310546875\\
600	1.9168974609375\\
650	1.91759716796875\\
700	1.915740234375\\
750	1.91591186523437\\
800	1.9164677734375\\
850	1.91839331054687\\
900	1.9178017578125\\
950	1.91780786132813\\
1000	1.91917456054688\\
};
\addlegendentry{Prop. CE + Prop. IMFC}

\end{axis}

\end{tikzpicture}%
}
 \caption{Effective throughput against maximum channel speed for a fixed \gls{PDR} of $30$ dB and \gls{SNR} of $15$ dB. The channel has integer delays and Doppler shifts and the transmitter adopts the $4$-QAM.} \label{fig:IntvsSpeed}
\end{figure}

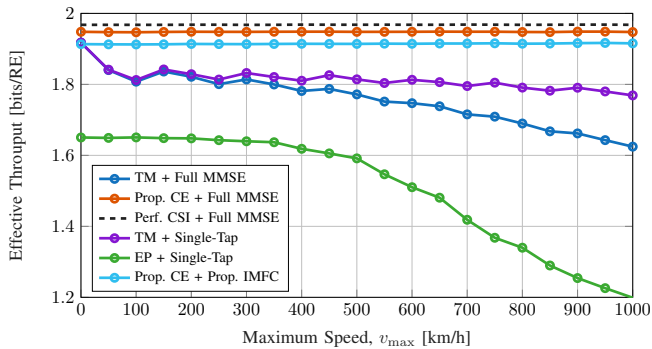
\begin{figure}[t]
 \centering
 \resizebox{0.99\columnwidth}{!}{
%
%
\definecolor{mycolor1}{rgb}{0.06600,0.44300,0.74500}%
\definecolor{mycolor2}{rgb}{0.86600,0.32900,0.00000}%
\definecolor{mycolor3}{rgb}{0.12941,0.12941,0.12941}%
\definecolor{mycolor4}{rgb}{0.52100,0.08600,0.81900}%
\definecolor{mycolor5}{rgb}{0.23100,0.66600,0.19600}%
\definecolor{mycolor6}{rgb}{0.18400,0.74500,0.93700}%
\begin{tikzpicture}

\begin{axis}[%
width=4.225in,
height=2.176in,
scale only axis,
xmin=0,
xmax=1000,
xticklabel style={
    /pgf/number format/set thousands separator={},
    /pgf/number format/1000 sep={}, 
},
xlabel style={font=\color{mycolor3}},
xlabel={Maximum Speed, $v_{\max}$ [km/h]},
ymin=1.2,
ymax=2,
ylabel style={font=\color{mycolor3}},
ylabel={Effective Throuput [bits/RE]},
axis background/.style={fill=white},
title style={font=\bfseries\color{mycolor3}},
xmajorgrids,
ymajorgrids,
legend style={at={(0.02,0.02)}, anchor=south west, legend cell align=left, align=left,font=\footnotesize,}
]
\addplot [color=mycolor1, line width=1.5pt, mark=o, mark options={solid, mycolor1}]
  table[row sep=crcr]{%
0	1.91761889648438\\
50	1.84026489257813\\
100	1.80768359375\\
150	1.83624584960937\\
200	1.82145239257812\\
250	1.80067041015625\\
300	1.81394653320313\\
350	1.79976318359375\\
400	1.78144873046875\\
450	1.78725073242187\\
500	1.77173388671875\\
550	1.7515498046875\\
600	1.74705615234375\\
650	1.73836206054688\\
700	1.71549951171875\\
750	1.70906274414062\\
800	1.68974169921875\\
850	1.66772827148438\\
900	1.66184252929688\\
950	1.64293798828125\\
1000	1.624587890625\\
};
\addlegendentry{TM + Full MMSE}

\addplot [color=mycolor2, line width=1.5pt, mark=o, mark options={solid, mycolor2}]
  table[row sep=crcr]{%
0	1.9481337890625\\
50	1.9471845703125\\
100	1.94682177734375\\
150	1.947716796875\\
200	1.94831079101562\\
250	1.9480478515625\\
300	1.94794384765625\\
350	1.94827001953125\\
400	1.94847045898438\\
450	1.94849365234375\\
500	1.94798168945312\\
550	1.94803979492188\\
600	1.94817016601562\\
650	1.94858520507812\\
700	1.94837109375\\
750	1.94839379882813\\
800	1.94757299804688\\
850	1.94728784179688\\
900	1.9486484375\\
950	1.94864819335938\\
1000	1.94751318359375\\
};
\addlegendentry{Prop. CE + Full MMSE}

\addplot [color=mycolor3, dashed, line width=1.5pt]
  table[row sep=crcr]{%
0	1.96777856445312\\
50	1.967796875\\
100	1.967767578125\\
150	1.96774340820313\\
200	1.96811108398437\\
250	1.96815087890625\\
300	1.96765893554687\\
350	1.96783374023438\\
400	1.96835888671875\\
450	1.96812622070312\\
500	1.96801904296875\\
550	1.96821752929688\\
600	1.9679892578125\\
650	1.96801391601562\\
700	1.96846557617188\\
750	1.9681962890625\\
800	1.96783911132813\\
850	1.96801147460938\\
900	1.96831909179687\\
950	1.96835473632813\\
1000	1.96785424804688\\
};
\addlegendentry{Perf. CSI + Full MMSE}

\addplot [color=mycolor4, line width=1.5pt, mark=o, mark options={solid, mycolor4}]
  table[row sep=crcr]{%
0	1.91758325195313\\
50	1.8414423828125\\
100	1.81230932617188\\
150	1.84202099609375\\
200	1.82871044921875\\
250	1.81357592773438\\
300	1.83203857421875\\
350	1.8207763671875\\
400	1.8102587890625\\
450	1.82594873046875\\
500	1.81417504882812\\
550	1.80386376953125\\
600	1.812916015625\\
650	1.80634448242188\\
700	1.79533935546875\\
750	1.80480810546875\\
800	1.79091650390625\\
850	1.78219848632813\\
900	1.79058764648438\\
950	1.78010815429688\\
1000	1.76897094726562\\
};
\addlegendentry{TM + Single-Tap}

\addplot [color=mycolor5, line width=1.5pt, mark=o, mark options={solid, mycolor5}]
  table[row sep=crcr]{%
0	1.6504\\
50	1.6492\\
100	1.6506\\
150	1.6485\\
200	1.6479\\
250	1.6428\\
300	1.6397\\
350	1.6370\\
400	1.6185\\
450	1.6055\\
500	1.5915\\
550	1.5467\\
600	1.5103\\
650	1.4807\\
700	1.4185\\
750	1.3678\\
800	1.3398\\
850	1.2897\\
900	1.2543\\
950	1.2259\\
1000	1.1988\\
};
\addlegendentry{EP + Single-Tap}

\addplot [color=mycolor6, line width=1.5pt, mark=o, mark options={solid, mycolor6}]
  table[row sep=crcr]{%
0	1.91341137695312\\
50	1.91279565429687\\
100	1.91254077148437\\
150	1.9128720703125\\
200	1.913923828125\\
250	1.91354614257813\\
300	1.9133193359375\\
350	1.91388549804687\\
400	1.91437231445313\\
450	1.91417553710937\\
500	1.91404345703125\\
550	1.9147646484375\\
600	1.91438793945312\\
650	1.91514111328125\\
700	1.91514672851562\\
750	1.91573999023438\\
800	1.91461352539062\\
850	1.91500805664062\\
900	1.9161953125\\
950	1.91677197265625\\
1000	1.9155810546875\\
};
\addlegendentry{Prop. CE + Prop. IMFC}

\end{axis}

\end{tikzpicture}%
}
 \caption{Effective throughput against maximum channel speed for a fixed \gls{PDR} of $30$ dB and \gls{SNR} of $15$ dB. The channel has fractional delays and Doppler shifts and the transmitter adopts the $4$-QAM.} \label{fig:FracvsSpeed}
\end{figure}

\subsection{Effective Throughput vs Speed}
To evaluate robustness under varying mobility conditions expected in 6G, Figures \ref{fig:IntvsSpeed} and \ref{fig:FracvsSpeed} illustrate the effective throughput as a function of maximum channel speed for integer and fractional channels, respectively. 
In Figure \ref{fig:IntvsSpeed}, the TM + Single-Tap method achieves near-optimal performance across all speeds, exhibiting only a minor throughput degradation even at very high mobility. The Prop. CE + Prop. IMFC method shows slightly lower throughput compared to TM + Single-Tap and Prop. CE + Full MMSE in this idealized integer channel scenario, due to the additional processing required to account for fractional effects that are absent here. 
Figure \ref{fig:FracvsSpeed} presents a more realistic scenario with fractional channel parameters. Here, the TM + Single-Tap method starts to fail even at moderate speeds. While it matches the throughput of Prop. CE + Prop. IMFC in static channels ($v_{\max}=0$), its performance rapidly degrades as the maximum speed increases due to high \gls{IPI} resulting from fractional delays and Doppler shifts. Conversely, the Prop. CE + Prop. IMFC maintains a nearly constant effective throughput across the entire speed range, demonstrating robust performance under high-mobility conditions. This highlights the suitability of the proposed method for practical high-speed \gls{OFDM} communications, where both fractional channel effects and \gls{ICI} are present.

\section{Conclusions and Future Works}
In this work, a novel receiver architecture for \gls{SISO} \gls{OFDM} high-mobility communications using delay-Doppler superimposed pilots has been proposed. The current literature on delay-Doppler \gls{SP} largely relies on the assumptions of integer delays and Doppler shifts, as well as \gls{ICI}-free operation. The proposed approach effectively addresses these limitations by: (i) enabling accurate estimation of fractional channel parameters, and (ii) incorporating \gls{ICI}-aware receiver processing. 

The receiver design maintains low complexity, leveraging disjoint delay-Doppler estimation and the Landweber-based \gls{IMFC} equalizer, whose complexity scales linearly with the number of \glspl{RE}. Extensive simulation results demonstrate that the proposed approach achieves nearly constant communication performance across a wide range of mobility scenarios, up to the maximum 6G speed of $1000$ km/h. These results highlight the potential of delay-Doppler \gls{SP} schemes to enhance \gls{OFDM} performance in future high-mobility 6G networks.

Future research directions include extending delay-Doppler \gls{SP} algorithms to multi-antenna \gls{OFDM} systems and exploring novel multiple superimposed pilot designs to reduce \gls{PAPR}, which is a key limitation of the current single-pilot approach. Additionally, given that the delay-Doppler domain is naturally suited for sensing applications, the performance of \gls{OFDM} with delay-Doppler \gls{SP} in \gls{ISAC} systems will be investigated.

\ifCLASSOPTIONcaptionsoff
  \newpage
\fi

\balance
\bibliographystyle{IEEEtran}
\bibliography{reference}

\end{document}